\documentclass[12pt]{article}
\pdfoutput=1
\usepackage{,amsmath,amsfonts,mathabx,amsbsy,latexsym,amssymb,amscd,amstext}
\usepackage{overpic,youngtab}
\usepackage{array}
\usepackage[scr=euler,cal=cm]{mathalfa}
\usepackage{color}
\usepackage[left=2.5cm,bottom=3cm,right=2.5cm,top=3cm]{geometry} 
\usepackage[latin1]{inputenc}
\usepackage{overpic,youngtab}
\usepackage{subfigure}
\usepackage[latin,english]{babel}
\usepackage{amsmath}
\usepackage{amssymb}
\usepackage{epstopdf}
\usepackage{graphics,psfrag,rotating}
\usepackage{mathabx}
\usepackage{graphicx}
\usepackage{dcolumn}
\usepackage{float}
\usepackage{pdflscape}
\usepackage{array}
\usepackage{booktabs}
\usepackage{amscd} 
\usepackage{mathtools}
\usepackage{fancybox}
\usepackage{fix-cm}
\usepackage[colorlinks=true,citecolor=blue,,linktocpage=true,linkcolor=blue,urlcolor=black]{hyperref}
\usepackage{tikz} 
\usetikzlibrary{matrix} 
\usetikzlibrary{positioning} 
\usetikzlibrary{arrows}
\usepackage{titlesec}
\usepackage{abstract}
\numberwithin{equation}{section}
\usepackage{cite}

\def\be{\begin{equation}}
\def\ee{\end{equation}}
\def\bea#1\eea{\begin{align}#1\end{align}}
\def\pd{\partial}
\def\a{\alpha}
\def\xp{x^\prime}
\def\b{\beta}

\def\g{\gamma}
\def\d{\delta}
\def\m{\mu}
\def\n{\nu}
\def\t{\tau}

\def\l{\lambda}

\def\r{\rho}

\def\bR{\bar{R}}

\def\s{\sigma}
\def\e{\epsilon}

\def\bma{\begin{pmatrix}}
	\def\ema{\end{pmatrix}}

\def\bg{\bar{g}}
\def\bi{\begin{itemize}}
	\def\ei{\end{itemize}}
\def\bn{\bar{\nabla}}

\begin{document}

	\vspace*{-1cm}
	\phantom{hep-ph/***} 
	{\flushleft
		{{FTUAM-20-9}}
		\hfill{{ IFT-UAM/CSIC-20-83}}}
	\vskip 1.5cm
	\begin{center}
		{\LARGE\bfseries Structural stability of spherical horizons}\\[3mm]
		\vskip .3cm
		
	\end{center}
	\vskip 0.5  cm
	\begin{center}
		{\large Enrique Alvarez, Jesus Anero  and Raquel Santos-Garcia }
		\\
		\vskip .7cm
		{
			Departamento de F\'isica Te\'orica and Instituto de F\'{\i}sica Te\'orica, 
			IFT-UAM/CSIC,\\
			Universidad Aut\'onoma de Madrid, Cantoblanco, 28049, Madrid, Spain\\
			\vskip .1cm

			\vskip .5cm
			\begin{minipage}[l]{.9\textwidth}
				\begin{center} 
					\textit{E-mail:} 
					{enrique.alvarez@uam.es, jesusanero@gmail.com, raquel.santosg@uam.es}
					
				\end{center}
			\end{minipage}
		}
	\end{center}
	\thispagestyle{empty}

\begin{abstract}
	\noindent
	This paper is concerned with the structural stability of spherical horizons. By this we mean stability with respect to variations of the second member of the corresponding differential equations, corresponding to the inclusion  of the contribution of  operators quadratic in curvature.  This we do both in the usual second order approach (in which the independent variable is the spacetime metric) and in the first order one (where the independent variables are the spacetime metric and the connection field). In  second order, it is claimed that the generic solution in the asymptotic regime (large radius) can be matched not only with the usual solutions with horizons (like Schwarzschild-de Sitter) but also with a more generic (in the sense that it depends on more arbitrary parameters) horizonless family of solutions. It is however remarkable that these horizonless solutions are absent in the {\em restricted} (that is, when the background connection is the metric one)  first order approach.
\end{abstract}

\newpage
\tableofcontents
\thispagestyle{empty}
\thispagestyle{empty}
\newpage
\setcounter{page}{1}

%\newpage
%%%%%%%%%%%%%%%

%\tableofcontents

\newpage

%%%%%%%%%%%%%%%%%%%%%%%%%%%%%%%%%%%%%%%%%%%%%%%%%%%%%%%%%%%%%%%%%%%%%%%%%%%%%%%%%%%%%%%%%%%%%%%%%%%%%%%%%%%%%%%%%%%%%%%%%%%%%%
\section{Introduction}
%%%%%%%%%%%%%%%%%%%%%%%%%%%%%%%%%%%%%%%%%%%%%%%%%%%%%%%%%%%%%%%%%%%%%%%%%%%%%%%%%%%%%%%%%%%%%%%%%%%%%%%%%%%%%%%%%%%%%%%%%%%%%%
The effective Lagrangian framework, whose origin goes back at least to Wilson's \cite{Wilson} pioneering work, dominates much of the research in particle physics in the last decades. It has been very successful, although the lack of new physics as yet in the LHC experiment is a clear indication that the related naturalness issue is still poorly understood.
\par
At any rate, when studying the gravitational interaction, and assuming diffeomorphism invariance (that is, general coordinate invariance or geometrization of gravity) as a fundamental symmetry \cite{AAGM}, the relevant operators in the infrared start with 
\be\label{action2}
S=\int d^4 x \sqrt{|g|} \, \Big[-\Lambda-\g R-2 \a R_{\m\n}^2+\left(\b+{2\over 3} \a \right)\,R^2+\ldots \Big],
\ee
where $\g\equiv {1\over 16\pi G}={1\over 2 \kappa^2}$, where $G$ is Newton's constant. The first two terms represent the Einstein-Hilbert Lagrangian with the cosmological constant. Incidentally, the smallness of the dimensionless combination $\Lambda\kappa^4$ is one of the facts that are not understood; we do not have anything new to say about it. 
\par
Regarding  the quadratic terms, in spacetime dimension $n=4$ (the only one considered in the present work), there are only two independent operators; we choose $R^2$ and $R_{\m\n}^2$ in order to agree with most of the existing literature. On the other hand, the appearance of terms quadratic (and higher) in curvature is unavoidable whenever the gravitational field is quantized in this geometric way unless a fine-tuning of sorts renormalizes all higher-order coefficients to zero.
\par
At this point, we have to decide which ones are our fundamental physical fields. In the usual {\em second order approach} (SO), the only such field is the metric tensor, and the connection is taken from the very beginning as the Levi-Civita one associated to the spacetime metric. Be that as it may, there is also the possibility of considering that {\em both the connection and the metric} are physical fields, with their own independent dynamics; this is the {\em first order approach} (FO). Curiously enough, when the Lagrangian is linear in curvature (Einstein-Hilbert) the equation of motion (EoM) of the connection forces it to be exactly the Levi-Civita one so that both approaches are equivalent \cite{Anero}, even when quantum corrections are considered. This  state of affairs does not hold true however for Lagrangians higher-order in curvature. There the connection field has a rich dynamics and it is not forced to be the Levi-Civita one\cite{AAGM,AlvarezGonzalez,AlvarezAneroSantos,AlvarezAneroSantos2}. 
\par
This means that we have to treat both approaches in turn in an independent way.
\par
We want to explore, in particular, the point of view that it is physically unavoidable to take all terms of higher-order in curvature seriously and explore their physical consequences (cf. \cite{Stelle78,Lu,Holdom,Holdom:2016nek,Holdom:2019ouz,Holdom:2020onl,Cano:2019ycn,Salvio:2019llz} and references therein). We would like to stress the fact that this is an issue independently of whether the ultraviolet (UV) completion of general relativity is to be based upon new (even non-geometric) variables, such as strings (cf. \cite{Alvarez} and references therein).
\par
Let us first explore the second order formalism, in which the connection is already explicitly taken as the Levi-Civita one associated to the metric tensor so that gravitation is fully represented by the said metric tensor. In \cite{Stelle78,Lu} they found among other things, a whole family of solutions unrelated to Schwarzschild's solution. In particular, the structural instability of the horizon under a particular type of quadratic corrections has also been pointed out previously by Holdom in \cite{Holdom,Holdom:2016nek}. 
% A pioneering work in our whole topic is the one of Holdom \cite{Holdom} who argued for the existence of nonsingular static and spherically symmetric solutions of the corrected equations. \textcolor{red}{Pensar cual de las dos frases quiero poner}
\par
This claimed instability is unrelated to the usual Regge-Wheeler stability \cite{Regge,Zerilli,Teukolsky} which is the ordinary stability keeping fixed the EoM. What we purport to do here is to change the EoM by a given amount (the quadratic corrections in the curvature), and analyze the structural stability of the solutions.
\par
%%%%%%%%%%%%%%%%%%%%%%%%%%%%%%%%%%%%%%%%%%%%%%%%%%%%%%%%%%%%%%%%%%%%%%%%%%%%%%%%%%%%%%%%%%%%%%%%%
Let us elaborate somewhat. Given an ordinary differential equation (ODE)
\be
\dot{x}=\xi(x),
\ee
where $x\in M$, and $M$ is some n-dimensional manifold, it is said that the system is {\em structurally stable} if it remains equivalent to itself when the vector field is changed by a small amount. All this can be made mathematically rigorous (see e.g. \cite{Arnold}). This concept was originally proposed by the soviet mathematicians Andronov and Pontriagin \cite{Andronov}. 
Let us be more precise. Andronov and Pontriagin considered a system
\bea
&{dx\over dt}=P(x,y),\nonumber\\
&{dy\over dt}=Q(x,y),
\eea
over a domain, $D\subset \mathbb{R}^2$. Andronov and Pontriagin  defined the system as structurally stable if for every $\e >0$ there exists a $\d>0$ such that whenever the functions $P$ and $Q$ are perturbed by other functions $p(x,y)$ and $q(x,y)$ smaller than $\d$, then the perturbed system
\bea
&{dx\over dt}=P(x,y)+p(x,y),\nonumber\\
&{dy\over dt}=Q(x,y)+q(x,y),
\eea
has trajectories that are displaced from those of the original system by less than $\e$. They were able to prove that this property is true if all singularities and closed orbits are hyperbolic and also there are no trajectories connecting saddles. Details can be consulted in  \cite{Pugh} and references therein.
It was later shown by 
Smale \cite{Smale} that in four dimensions there is a vector field that cannot be made structurally stable by a small deformation.
\par
Although these ideas are restricted to ODE (and actually most useful in the original two-dimensional setting), the general concept can be applied in a somewhat loose sense to partial differential equations (PDE) ({\em confer} the use of this concept in \cite{Paquette}); it is in this generalized way that we are using this concept here, without any pretension of mathematical rigor whatsoever. 
\par
Let us briefly review the ideas behind the setup that we will use throughout the paper. Once the conditions for spherical symmetry have been used, the spacetime metric has been assumed to take the form
\be
ds^2= B(r) dt^2-A(r) dr^2-r^2 d\Omega_2^2.\label{metricansatz}
\ee
\par
The idea is to change the second member of equations \eqref{metricansatz} with terms implied by a generic Lagrangian quadratic in curvature. To begin with, we shall keep the assumption of spherical symmetry and staticity. We are well aware that this does not correspond to a generic perturbation, but of course, any eventual instability found in this particular case is bound to survive in the general situation. 
\par
Let us now summarize the main findings of \cite{Stelle78,Lu}. They classify the solutions by the behavior of the functions appearing in the metric \eqref{metricansatz}  at $r\sim 0$, where they expand them as
\bea
&A(r)=a_s r^s+a_{s+1}r^{s+1}+\ldots\nonumber\\
&B(r)=b_t\left(r^t+b_{t+1}r^{t+1}+\ldots\right),
\eea
The behavior of the different solutions is summarized in the following
\bi
\item There is a 3-parameter family with the behavior $(s,t)=(0,0)$. Those solutions are not singular and correspond to candidates for the  {\em vacuum} of the theory.
\item There is another 4-parameter singular family with the behavior $(s,t)=(1,-1)$. Schwarzschild's space-time falls in this category. In this same reference, it was argued that this family cannot be coupled to a normal (codimension 1, shell) source.
\item There is a 6-parameter singular family with the behavior $(s,t)=(2,2)$. These solutions have no horizon and are the only ones that can be coupled to normal matter.
\ei
\par
The main purpose of the present work is to examine the possibility that the six-parameter family of solutions uncovered by \cite{Stelle78,Lu} (or rather its generalization with cosmological constant) can be an alternative approximation to the gravitational field of a static, spherical source, different close to the origin to Schwarzschild's spacetime itself. Our idea is that all Solar system gravitational tests are in fact done in the region of the metric corresponding to  ${r\over r_s}\gg 1$ (although $\frac{r}{r_\Lambda}\ll 1$). Here the Schwarzschild's radius is given by $r_s\equiv  {\kappa^2 M_\odot\over 4\pi}$ and the length scale associated to the cosmological constant is given by $r_\Lambda\equiv \sqrt{6 \g\over \Lambda}\sim{1\over \kappa^3}$;  this  is in the (intermediate) asymptotic region.  We lack any Solar system test of the region $r\sim r_s$. It is our purpose to check explicitly whether it is possible to analytically match different solutions (in particular the horizonless $(2,2)$ family and the Schwarzschild-de Sitter type $(1,-1)$ one) with the asymptotic behavior of the solution. % As we will see, the asymptotic behavior of the solution (that is, in the region ${r\over r_s}\gg 1$) is universal. In the case of $\Lambda =0$ the metric functions behave as
We already know that it is possible to do so with the Schwarzschild-de Sitter  solution, which is still a solution of the modified EoM. When working with the Einstein-Hilbert Lagrangian,  any static spherically symmetric solution is isomorphic \cite{Schleich} to (a region of) the  Schwarzschild-de Sitter or the related Nariai spacetimes.  For quadratic theories however, the solution is not unique anymore and some of the arguments for the issue of the final state after the total collapse are questionable. 
\par
While carrying out this analysis we also hope to clarify a few points regarding the runaway type of solutions in \cite{Lu,Holdom:2016nek}. We comment in particular in Appendix \ref{appB} on the existence of these runaway solutions, which is a generic fact of higher-order equations. Despite many efforts by many authors (mainly in the context of the Lorentz-Dirac equation), there is no accepted and systematic way of eliminating them. They are in some sense the classical counterpart of the existence of ghosts in the quantum version of the theory.
\par
Let us now turn to the FO approach. We have already commented upon the fact that both approaches are not equivalent in general once operators of higher-order in curvature are included in the Lagrangian. Actually, kinematically, the connection field embodies one spin 3 component, a set of three
spin 2 components, five spin 1 components and three spin 0 components \cite{AlvarezAneroSantos}. Nonetheless, a fully general analysis is too involved to be included here.\par
It is remarkable that even when the background connection field is the Levi-Civita one, FO (we shall dub it {\em restricted} FO then) and SO are still not totally equivalent. We shall prove that every FO solution is also a solution in SO, but the converse is not true. What is most remarkable is that this is precisely what happens with those horizonless solutions: they are absent in the restricted FO approach. 
\par
The plan of the paper is as follows\footnote{
	Throughout  this work we follow the Landau-Lifshitz spacelike conventions, in particular our metric has the signature $\eta_{\m\n}=(+,-,-,-)$ and $R^\m_{~\n\r\s}=\partial_\r \Gamma^\m_{\n\s}-\partial_\s \Gamma^\m_{\n\r}+\Gamma^\m_{\l\r}\Gamma^\l_{\n\s}-\Gamma^\m_{\l\s}\Gamma^\l_{\n\r}$. 
	%we define the Ricci tensor as
	%\be R_{\m\n}\equiv R^\l_{~\m\l\n}\ee
	%and the commutator with our conventions is
	%\bea[\bn_\m,\bn_\n]V^\l_\t&=\bR^\l_{~\r\m\n}V^\r_\t-\bR^\r_{~\t\m\n}V^\l_\r\label{c}\eea
}.
First, in section 2, we explain the background results, which we will use in the rest of the paper. Section 3 is devoted to the generalization of the analysis in \cite{Lu,Holdom,Holdom:2016nek} in the presence of a cosmological constant to analyze whether any of the newly found (albeit in an incomplete form) solutions are physically acceptable. 
%We also study a particular case of quadratic Lagrangians which can be written as a sum of squares, where the $(2,2)$ family is no longer a solution. 
In section 4, we expand the solutions around an arbitrary intermediate point $r_0$. We show that the solutions around this arbitrary point have enough parameters to match them with the families of solutions $(2,2)$ or $(1,-1)$ that appear near the origin.
\par 
In the second part of this work, after section 5, we elaborate on the FO approach. After some general comments in section 5, it is argued that it would be simpler to consider only the restricted FO (that is, assume the connection field to be Levi-Civita one). We have already pointed out that even in this case FO is not strictly equivalent to SO; there are SO solutions that are not FO solutions. This will be discussed in some detail. This restricted procedure is however not helpful when the background metric is assumed to be Ricci flat or for constant curvature spacetimes, because then the EoM for the graviton (the metric perturbations) is tautological; it does not give any information. There we can assume that the spacetime metric comes from one of the spin 2 fields hidden in the connection field. In section 5, we work out some simple examples for maximally symmetric spaces to illustrate the formalism.  In section 6, we carry out the same study for the Schwarzschild background solution. When the background is not Ricci flat, as in the most important case of Schwarzschild-de Sitter, we can restrict the connection to the Levi-Civita form, and work out the graviton EoM, which is now not empty. In section 7, we turn again to the study of possible spherically symmetric solutions in the restricted FO approach, carrying out the same analysis as in section 2. In this case, the restricted FO approach does not admit the horizonless power series solutions found in the SO approach. We find this fact one of the most important results of our work. Finally, we present our conclusions and remarks. Given the length of some of the equations involved in the computation, we relegate most of them to the appendices.
%%%%%%%%%%%%%%%%%%%%%%%%%%%%%%%%%%%%%%%%%%%%%%%%%%%%%%%%%%%%%%%%%%%%%%%%%%%%%%%%%%%%%%%%%%%%%%%%%%%%%%%%%%%%%%%%%%%%%%%%%%%%%%%
\section{Preliminaries}
%%%%%%%%%%%%%%%%%%%%%%%%%%%%%%%%%%%%%%%%%%%%%%%%%%%%%%%%%%%%%%%%%%%%%%%%%%%%%%%%%%%%%%%%%%%%%%%%%%%%%%%%%%%%%%%%%%%%%%%%%%%%%%%
\bi
\item
Let us imagine that the only thing known is Schwarzschild's solution 
\be
ds^2=\left(1-\frac{r_s}{r}\right)dt^2-\frac{1}{\left(1-\frac{r_s}{r}\right)}dr^2-r^2 d\Omega_2^2,
\ee
where $r_s=2GM$, is the expansion of the metric functions in \eqref{metricansatz} around $r=0$ as 
\be
\left.B(r)\right|_{r\sim 0}={r_s\over r}\left(-1+{r\over r_s}\right)+\ldots 
\ee
We know from the exact solution that this expansion stops there. The expansion of the other function behaves as 
\be
\left.A(r)\right|_{r\sim 0}=-{r\over r_s}\left(1+{r\over r_s}\right)+\ldots\,
\ee
and the exact expansion here is given by
\be
\left.A(r)\right|_{r\sim 0}=-{r\over r_s}\sum_{n=0}^\infty\left({r\over r_s}\right)^n.
\ee

In the opposite limit, its asymptotic behavior when $r\rightarrow\infty$,  the  function reads
\be
\left.B(r)\right|_{r\sim \infty}=1-{r_s\over r}+\ldots
\ee
and the full expansion also stops here. For the other function we have 
\be
\left.A(r)\right|_{r\sim \infty}=1+{r_s\over r}+\ldots
\ee
and again, the complete expansion is known to be
\be
\left.A(r)\right|_{r\sim \infty}=\sum_{n=0}^\infty\left({r_s\over r}\right)^n.
\ee

It is also possible to  expand the metric functions around an arbitrary point $r=r_0$ 
\bea B(r)&=1-\frac{r_s}{r_0}\sum_{n=0}^{\infty}\left(\frac{r_0-r}{r_0}\right)^n,\nonumber\\
A(r)&=\frac{r_0}{r_0-r_s}+\frac{r_s}{r_0-r_s}\sum_{n=1}^{\infty}\left(\frac{r_0-r}{r_0-r_s}\right)^n.\eea
However, in a perturbative computation of these functions through some ODE, one  would not in general have access to the general term of the series. 
\par

In case we knew the full expansion it would allow for immediate access to the radius of convergence $\r$ of the power series, $\sum a_n r^n$, through e.g. Hadamard 's criterium
\be
{1\over \r}\equiv \lim\limits_{n\rightarrow\infty}\,|a_n|^{1\over n}.
\ee
This yields immediately when expanding around the origin, $\r=r_s$, for the radius of convergence of our series. 
%When expanding around $r=\infty$  we get
%\be
%{r_s\over r}\leq 1.
%\ee
Nevertheless, even without exact knowledge,  knowing just a few terms in the expansion gives some estimate of the radius of convergence. 
\par
In the example above  we would tentatively estimate the correct radius of convergence as
\be
{1\over \r}\sim \{a_1,\,a_2^{1\over 2},\,a_3^{1\over 3},\,\ldots\}.
\ee

\item
The matching between two power series can be done when both power series are the analytic continuation one of the other. For example,
\be\label{uno}
\sum_{n=0}^\infty r^n,
\ee
which converges for $r\leq 1$, can be matched with
\be\label{dos}
{1\over 1-r_0}\sum_{n=0}^\infty \left({r-r_0\over 1-r_0}\right)^n,
\ee
which converges whenever $\left|r-r_0\right|\leq \left|1-r_0\right|$.
\par
Again, it is difficult to make precise statements without knowing the general term of the series; but an evident necessary condition for it to be possible is that there are enough free parameters to do so (in this example, the first series \eqref{uno} matches with the second \eqref{dos} when $r_0=0$).

\item
On the other hand, when we are interested, as in the case at hand, in solutions of a certain system of ordinary differential equations (ODE), there is a useful theorem.
Given such a set \footnote{Please note that {\em every} ODE system of arbitrary order can easily be written in this first order form.
} 
\be
{d y_i\over dx}=f_i\left(x,\,y_1,\ldots\, y_n\right)=\sum_{q,\,q_i\geq 0} C^i_{p\, q_1\ldots q_n} x^p\, y_1^{q_1}\,\ldots\, y_n^{q_n},
\ee
where it is assumed that the power series in the second member  converges for
\be
|x|,|y_i|\leq \r_s 
\ee 
Denote by  $M$ the least upper bound of the functions $f_i(x,y)$ in the set as above (of course, in most cases this is not known until the full solution is found). Then 
\be
C^i_{p,q}\leq {M\over \r_s^{p+q_1+\ldots+ q_n}}.
\ee
A theorem exists \cite{Cartan} ensuring that the solution to the ODE system has a radius of convergence such that
\be
\r=\r_s\left(1-e^{-{1\over (n+1)M}}\right).
\ee
It follows that when the second member of the ODE is not bounded (that is $M=\infty$) then $\r=0$.
\item
In the spherically symmetric case, it is well known that Einstein's equations are equivalent to the following system of equations in terms of the metric functions
\be
A={1\over B},
\ee
\be
{d A\over dr}={A(1-A)\over r}\equiv f(r,A).
\ee
The series expansion of the function on the rhs then reads
\be
f(r,A)={1\over r_0}\sum_{n=0}^\infty \left({r-r_0\over r_0}\right)^n A\left(1-A\right),
\ee
so that
\be
{1\over \r_s}\equiv  \lim\limits_{n\rightarrow\infty}\,{1\over r_0^{n+1\over n}}={1\over r_0},
\ee
and
\be
\r=r_0\left(1-e^{-{1\over 2 M}}\right),
\ee
where $M$ is defined as
\be
M\equiv \lim\limits_{r\leq r_0} A,
\ee 
which as advertised, is not known {\em a priori}.
\ei
%%%%%%%%%%%%%%%%%%%%%%%%%%%%%%%%%%%%%%%%%%%%%%%%%%%%%%%%%%%%%%%%%%%%%%%%%%%%%%%%%%%%%%%%%%%%%%%%%%%%%%%%%%%%%%%%%%%%%%%%%%%%%%%
\section{New solutions in the second order approach}
%%%%%%%%%%%%%%%%%%%%%%%%%%%%%%%%%%%%%%%%%%%%%%%%%%%%%%%%%%%%%%%%%%%%%%%%%%%%%%%%%%%%%%%%%%%%%%%%%%%%%%%%%%%%%%%%%%%%%%%%%%%%%%
Let us generalize  the analysis carried out in \cite{Lu} in the presence of a  cosmological constant. The EoM for the action, \eqref{action2}, reads
\bea &H^{SO}_{\m\n}=\g\left(R_{\m\n}-\frac{1}{2}g_{\m\n}R\right)-2\left(\b+\frac{2\a}{3}\right)R\left(R_{\m\n}-\frac{1}{4}g_{\m\n}R\right)+4\a R^{\r\s}\left(R_{\m\r\n\s}-\frac{1}{4}g_{\m\n}R_{\r\s}\right)+\nonumber\\
&+\frac{2}{3}\left(\a-3\b\right)\left(g_{\m\n}{\Box}-\nabla_\m\nabla_\n\right)R+2\a \bar{\Box}\left(R_{\m\n}-\frac{1}{2}g_{\m\n}R\right)-\frac{1}{2}g_{\m\n}\Lambda = 0\label{EOMB}.
\eea
Schwarzschild-de Sitter's metric is given by
\be
ds^2=\left[1-\frac{r_s}{r}-\left(\frac{r}{r_\Lambda}\right)^2\right]dt^2-\frac{1}{\left[1-\frac{r_s}{r}-\left(\frac{r}{r_\Lambda}\right)^2\right]}dr^2-r^2\left(d\theta^2+\sin^2{\theta}d\phi^2\right),
\label{schwdS}
\ee
where $r_s=2GM$ is the Schwarzschild radius, and  $r_\Lambda^2=\frac{6\g}{\Lambda}$, and is a particular and in some sense archetypal solution of the EoM.
%%%%%%%%%%%%%%%%%%%%%%%%%%%%%%%%%%%%%%%%%%%%%%%%%%%%%%%%%%%%%%
\par
Solutions to quadratic gravity of the Petrov types $N$ and $III$ have been recently studied by Malek and Pradva \cite{Malek:2011pn}. In fact  R.~Svarc, J.~Podolsky, V.~Pravda and A.~Pravdova \cite{Svarc:2018coe, Pravda:2020zno} were able to show that Einstein spaces are always solutions of a generic quadratic gravity EoM, barring some algebraic connection between the cosmological constant and the coupling constants of the lagrangian.

 \par
 Tha aim of our paper is to  search for more general solutions, assuming  only spherical symmetry and staticity.

%%%%%%%%%%%%%%%%%%%%%%%%%%%%%%%%%%%%%%%%%%%%%%%%%%%%%%%%%%%%%%%
\par
We want to analyze the possible families of solutions depending on the behavior of the functions appearing in the ansatz of the spherically symmetric metric \eqref{metricansatz}. Near the origin $r=0$ assuming the functions $A(r)$ and  $B(r)$ admit a Frobenius-like expansion
\bea
&A(r)=a_s r^s+a_{s+1}r^{s+1}+\ldots\nonumber\\
&B(r)=b_t\left(r^t+b_{t+1}r^{t+1}+\ldots\right),
\label{frobenius}
\eea
we recover {\em exactly} the same families as when $\Lambda=0$ \cite{Stelle78,Lu}, depending on the same number of parameters (besides the cosmological constant). In our case, we hereinafter ignore  (that is, we will not count it as a relevant parameter) the constant $b_t$ whose physical meaning is a change of the origin of the time coordinate, although it is counted as a parameter in \cite{Lu}. 
\par
The curvature scalar is given by
\be R=\frac{1}{2r^2AB}\Big\{-4A^2B^2-rBA'\left(4B+rB'\right)+A\left[4B^2-r^2B'^2+2rB\left(2B'+rB^{''}\right)\right]\Big\}\ee
In terms of the formal series expansion this reduces to 

\be R=\frac{r^{-2-s}}{2a_s}\left[4+2t+t^2-s(4+t)-4r^sa_s\right]\ee

\par
We continue with a more general case. Let us summarize our results.
\par
After using Bianchi's identities ($\nabla^\m H^{SO}_{\m\n}=0$) on the EoM \eqref{EOMB}, we find that only two of them are independent. We chose them as follows
\be
H^{SO}_{00}=H^{SO}_{11}=0.
\ee
With the ansatz \eqref{metricansatz} of the metric,  we can study different solutions of the EoM as a function of the $A(r)$ and $B(r)$ that will be solutions of 
\bea \label{EOMC}
&H^{SO}_{00}=\dfrac{1}{24 r^4A^5 B^3} \left[-24A^4 B^4(12\b+r^2\g)-4 A^5B^4\left(4\a-12\b-6r^2\g+3r^4\Lambda\right)+ \right. \nonumber\\
&+8A^3B^4(2\a+30\b)-16r^4 A^3 B^3(\a-3\b)B^{(4)}-4r^2A^2 B^2\left[3r(4\a-30\b)A'+(5\a-12\b)A\right]B'^2+\nonumber\\
&+16r^3A^2 B^2(\a-3\b)\left[3rA' B+3rA B'-4AB\right]B^{(3)}+56r^3B^3\left[-2(\a+6\b)B+r(\a-3\b)B'\right]A'^3+\nonumber\\
&+8r^3A^2 B^3\left[-2(\a+6\b)B+r(\a-3\b)B'\right]A^{(3)}+52r^3A B^3\left[2(\a+6\b)B-r(\a-3\b)B'\right] A'A''+\nonumber\\
&+8r^2 A^2 B^2\left[-3r^2(\a-3\b)B'^2-2(\a+6\b)B^2+4r^2(\a-3\b)BB''+3r(\a-6\b)B B'\right]A''+\nonumber\\
&+4r^3 A B^2\left[-19r(\a-3\b)A' B-27r(\a-3\b)A B'+2(13\a-48\b)AB\right]A' B''+\nonumber\\
&+36r^4(\a-3\b)A^3 B^2B''^2+4r^3A^3B \left[-29r(\a-3\b)B'+2(13\a-66\b)B\right]B' B''+\nonumber\\
&+4r^2AB^4(7\a+60\b)A'^2+r^3AB^2\left[57r(\a-3\b)B'-4(13\a-84\b)B\right]B'A'^2+\nonumber\\
&+r^3A^2\left[58r(\a-3\b)A'B+49r(\a-3\b)AB'-4(11\a-78\b)AB\right]B'^3+\nonumber\\
& \left. +8rA^2B^4\left[4(\a+6\b)+3r^2\g A\right]A'+8r^2A^2B^3A'B'(-5\a+6\b)\right]= 0,
\eea 
\bea
\label{EOMD}
&H^{SO}_{11}= \dfrac{1}{24 r^4A^3 B^4} \left[8r^3A^2B^2B^{(3)}\left[r(\a-3\b)B'-2(\a+6\b)B\right]-4r^4(\a-3\b)A^2B^2 B''^2- \right. \nonumber\\
&-4r^2ABB''\Big[2rBA'\left(r(\a-3\b)B'-2(\a+6\b)B\right)+A\left(3r^2(\a-3\b)B'^2-12r(\a+3\b)BB'+\right.\nonumber\\
&\left.+8(\a+6\b)B^2\right)\Big]+24A^3B^3\left[\g r^3B'+B(\g r^2-12\b)\right]+4A^4 B^4\left[4\a-12\b-6\g r^2+3r^4\Lambda\right]+\nonumber\\
&+2r^2ABA'B'\left[3r^2(\a-3\b)B'^2-4r(2\a+3\b)BB'+4(\a+24\b)B^2\right]+A^2\Big[7r^4(\a-3\b)B'^4-\nonumber\\
&-4r^3(5\a+12\b)BB'^3-4r^2(\a-48\b)B^2B'^2+32r(\a+6\b)B^3B'-16(\a-21\b)B^4\Big]\nonumber\\
&+7r^2B^2A'^2\left[r^2(\a-3\b)B'^2-4r(\a+6\b)BB'+4(\a-12\b)B^2\right]-4r^2AB^2A''\left[r^2(\a-3\b)B'^2-\right. \nonumber\\
&\left.\left.-4r(\a+6\b)BB'+4(\a-12\b)B^2\right]  \right]=0,
\eea
where $A^{(n)}$ corresponds to the n-th partial derivative. These equations coincide with the ones in \cite{Lu} when taking $\Lambda=0$.
%%%%%%%%%%%%%%%%%%%%%%%%%%%%%%%%%%%%%%%%%%%%%%%%%%%%%%%%%%%%%%%%%%%%%%%%%%%%%%%%%%%%%%%%%%%%%%%%%%%%%%%%%%%%%%%%
\subsection{Solutions around $r=0$}\label{r=0}
%%%%%%%%%%%%%%%%%%%%%%%%%%%%%%%%%%%%%%%%%%%%%%%%%%%%%%%%%%%%%%%%%%%%%%%%%%%%%%%%%%%%%%%%%%%%%%%%%%%%%%%%%%%%%%%%

We start by analyzing the solutions around the origin $r=0$. Around this point we take the expansions \eqref{frobenius}.
%\bea
%&A(r)=a_s r^s+a_{s+1}r^{s+1}+\ldots\nonumber\\
%&B(r)=b_t\left(r^t+b_{t+1}r^{t+1}+\ldots\right).
%\eea
It is perhaps worth remarking that owing to the fact that the functions $A$ and $B$ are dimensionless, each of the independent parameters in the power series defines a length scale given by
\bea
&a_i\equiv \left({1\over r^a_{(i)}}\right)^i, \quad \Lambda\equiv {6\g \over r_\Lambda^2}\nonumber\\
&b_t\equiv \left({1\over r^b_{(t)}}\right)^t,\quad b_j\equiv \left({1\over r^b_{(j)}}\right)^{j-t}\quad (j>t). 
\eea
We will sometimes denote the scale associated with the first non-vanishing coefficient in the power series just by $r_a$ or $r_b$ with no further subscript.
\par
Let us classify then the solutions in terms of the values of $(s,t)$.
\bi 
\item
The $(0,0)$ family of solutions takes the form 
%%%%%%%%%%%%%%%%%%%%%%%%%%%%%%%%%%%%%%%%%%%%%%%%%%%%%%%%%%%%%%%%%%%%%%%%%%%%%%%%%%%%%%%%%%%%%%%%%%%%%%%%%%%%%%%%
\bea &A(r)=1+a_2 r^2+r^4\Big\{\frac{\Lambda(-4\a+3\b)}{1080\a\b}+ \dfrac{1}{180 \a \b} \left[a_2[\g(2\a+3\b)+18\b a_2(10\a+3\b)] \right. \nonumber\\
&+ \left. b_2[2\g(-\a+3\b)-18\b b_2(2\a+3\b)]-36\a\b a_2 b_2 \right]+O(r^5)\nonumber\\&&\nonumber\\
&\frac{B(r)}{b_0}=1+b_2 r^2+r^4\Big\{\frac{\Lambda(2\a+3\b)}{2160\a\b}+\dfrac{1}{360 \a \b} \left[ a_2[\g(-\a+3\b)+54\b^2 a_2]+b_2[\g(\a+6\b) \right. \nonumber\\
& \left. +\frac{+54\b b_2(2\a-\b)]+108\a\b a_2 b_2}{360\a\b}\right]+O(r^5)
\label{00orig}
\eea
These solutions depend on three parameters $( a_2, b_2, \Lambda)$, and we recover the result in  \cite{Stelle78,Lu} when $\Lambda=0$ (we do not count the trivial time rescaling parameter $b_{0}$). 
\par
Our rough estimate for the radius of convergence of the power series is
\bea
&\rho (A)\sim r_a,\nonumber\\
&\rho (B)\sim r_b.
\label{radiusconv}
\eea
%%%%%%%%%%%%%%%%%%%%%%%%%%%%%%%%%%%%%%%%%%%%%%%%%%%%%%%%%%%%%%%%%%%%%%%%%%%%%%%%%%%%%%%%%%%%%%%%%%%%%%%%%%%%%%%%
\item
The second family of solutions is the $(1,-1)$ for which the functions read
%%%%%%%%%%%%%%%%%%%%%%%%%%%%%%%%%%%%%%%%%%%%%%%%%%%%%%%%%%%%%%%%%%%%%%%%%%%%%%%%%%%%%%%%%%%%%%%%%%%%%%%%%%%%%%%%
\bea 
&A(r)=a_1 r-a_1^2 r^2+a_1^3 r^3+a_4 r^4-r^5\frac{a_1}{16}\left(3a_1 b_2+19a_1^4+35a_4\right)+\nonumber\\
&+r^6\frac{a_1^2}{40}\left(21a_1 b_2+101a_1^4+141a_4\right) + O(r^7)  \nonumber \\
\label{Aorigin}
%+r^7\Bigg\{\frac{(-26 \a + 33 \b) \Lambda 
%    a_1^3}{11664\a\b}+\frac{1}{77760 \a\b a_1} \Big[30\g (13 \a + 33 \b)  
%a_1^6-\nonumber\\
%&& 
%-81 \b(2923 \a - 165 \b) a_1^8
%- 
%162\b (1489 \a - 165 \b)  
%a_1^4 a_4 + 
%405 \b (181 \a + 33 \b) 
%a_4^2+\nonumber\\
%&&+ 
%10\g (-65 \a + 231 \b)  
%a_1^3 b_2- 
%54 \b (1967\a+ 165 \b) 
%a_1^5 b_2 - 
%270 \b(109 \a + 33\b)a_1 a_4 b_2+\nonumber\\
%&&+ 15 
%a_1^2\left[\g(26 \a 
%+ 66 \b)a_4- 
%45 \b (37 \a+ 33 \b) 
%b_2^2\right]\Big]\Bigg\}+O(r^8)\nonumber\\&&\nonumber\\
\\
&\frac{B(r)}{b_{-1}}=\frac{1}{r}+a_1+b_2 r^2+ r^3\frac{1}{16}\left(a_1 b_2+a_1^4+a_4\right)-r^4\frac{3a_1}{40}\left(a_1 b_2+a_1^4+a_4\right)+\nonumber\\
&+r^5\Bigg\{\frac{(2 \a + 3 \b) \Lambda
	a_1^2}{3888\a\b}+\frac{1}{25920\a \b
	a_1^2} \Big[-30\g (\a - 
3 \b)  a_1^6
+ 
81\b(19 \a + 15 \b) 
a_1^8+\nonumber\\
&+ 
162 \b(7 \a + 15 \b) 
a_1^4 a_4- 
405\b (\a - 3 \b) 
a_4^2 + 
10\g (5 \a + 21 \b)
a_1^3 b_2 + 
54\b (161 \a - 15\b) 
a_1^5 b_2 + 
\nonumber\\
&+270 \b(25\a - 3 \b)a_1 a_4 b_2 - 15 
a_1^2 \left[2 \g(\a- 
3 \b) a_4 + 
9 \b (-53 \a+ 15\b) 
b_2^2\right]\Big]\Bigg\}+O(r^6)
\label{Borigin}
\eea
These solutions depend on four parameters $( b_2, a_1, a_4, \Lambda)$ and when $\Lambda=0$ we recover again the family of solutions in \cite{Stelle78,Lu}. 
The expansion of the Schwarzschild-de Sitter metric around the origin belongs to this family of solutions. Let us note that the cosmological constant appears at $O(r^7)$, although it is not written in order to avoid too long equations. The radius of convergence is the same as the previous family. 
%%%%%%%%%%%%%%%%%%%%%%%%%%%%%%%%%%%%%%%%%%%%%%%%%%%%%%%%%%%%%%%%%%%%%%%%%%%%%%%%%%%%%%%%%%%%%%%%%%%%%%%%%%%%%%%%
\item 
Finally, we have the $(2,2)$ family of solutions which has the form
%%%%%%%%%%%%%%%%%%%%%%%%%%%%%%%%%%%%%%%%%%%%%%%%%%%%%%%%%%%%%%%%%%%%%%%%%%%%%%%%%%%%%%%%%%%%%%%%%%%%%%%%%%%%%%%%
\bea &A(r)=a_2 r^2+a_2b_3 r^3-r^4\frac{a_2}{6}\left(2a_2+b_3^2-8b_4\right)+a_5 r^5+\nonumber\\
&+\frac{r^6}{1296\a\b}\Bigg\{-12\a^2a_2^3-2a_2^2\left[b_3^2\left(\a^2-603\a\b-252\b^2\right)+27\a\left(20\b b_4+\g\right)\right]+\nonumber\\
&+a_2\Big[b_3^4\left(-16\a^2+1413\a\b-72\b^2\right)+2b_4b_3^2\left(19\a^2-2223\a\b+180\b^2\right)-\nonumber\\
&-36b_5 b_3\left(\a^2+45\b^2\right)+12\a b_4^2\left(\a+162\b\right)\Big]+324a_5\b b_3\left(7\a+3\b\right)\Bigg\}+ O(r^7)\nonumber\\
\nonumber \\
&\frac{B(r)}{b_2}=r^2+b_3 r^3+b_4 r^4+b_5 r^5+\frac{r^6}{216\a a_2}\Bigg\{-12\a a_2^3+a_2^2\left[14b_3^2(2\a+3\b)-24\a b_4\right]+\nonumber\\
&+a_2\Big[2b_3^4(67\a-3\b)+2b_4 b_3^2(15\b-227\a)+45b_5 b_3(7\a-3\b)+180\a b_4^2\Big]\nonumber\\
& +27a_5b_3(\a+3\b)\Bigg\}+ O(r^7)
\eea
These solutions depend on the six parameters $(b_3, b_4, b_5, a_2, a_5,\Lambda)$ and again when $\Lambda=0$ it reduces to  the result in \cite{Stelle78}. Besides, these solutions are exactly the same as the ones in \cite{Stelle78,Lu} up to order $O(r^{10})$ where the terms containing the cosmological constant appear. The radius of convergence is again the one given in \eqref{radiusconv}.
%%%%%%%%%%%%%%%%%%%%%%%%%%%%%%%%%%%%%%%%%%%%%%%%%%%%%%%%%%%%%%%%%%%%%%%%%%%%%%%%%%%%%%%%%%%%%%%%%%%%%%%%%%%%%%%%
%\item \textcolor{red}{La ponemos? no aporta nada}
%The $(-2,2)$ family.
%%%%%%%%%%%%%%%%%%%%%%%%%%%%%%%%%%%%%%%%%%%%%%%%%%%%%%%%%%%%%%%%%%%%%%%%%%%%%%%%%%%%%%%%%%%%%%%%%%%%%%%%%%%%%%%%%
%This case is not a real solution.In fact
%\bea
%A(r)=a_{-2} r^{-2}+a_{-1}r^{-1}+\ldots\nonumber\\
%B(r)=b_2\left(r^2+b_{3}r^{3}+\ldots\right)
%\eea
%we obtain $a_{-2}=0$ or $b_2=0$, then we conclude this a spurious solution.
\ei

%%%%%%%%%%%%%%%%%%%%%%%%%%%%%%%%%%%%%%%%%%%%%%%%%%%%%%%%%%%%%%%%%%%%%%%%%%%%%%%%%%%%%%%%%%%%%%%%%%%%%%%%%%%%%%%%%%%%%%%%%%%%%%
\subsection{Asymptotic expansions}\label{rinf}
%%%%%%%%%%%%%%%%%%%%%%%%%%%%%%%%%%%%%%%%%%%%%%%%%%%%%%%%%%%%%%%%%%%%%%%%%%%%%%%%%%%%%%%%%%%%%%%%%%%%%%%%%%%%%%%%
When the cosmological constant is different from zero, we no longer have asympotically flat solutions. In this case we obtain a  
%The best we can do is to behave as 
de Sitter ($dS$) or anti-de Sitter space ($AdS$) behavior asymptotically, given by the expansion of the metric in \eqref{schwdS}.
%that is, the mric is given by
%\be
%ds^2=\left[1-\frac{r_s}{r}-\left(\frac{r}{r_\Lambda}\right)^2\right]dt^2-\frac{1}{\left[1-\frac{r_s}{r}-\left(\frac{r}{r_\Lambda}\right)^2\right]}dr^2-r^2\left(d\theta^2+\sin^2{\theta}d\phi^2\right).
%\ee
\par
Given the two scales that appear in the solution, there are actually two regimes that are interesting to study.
\bi
\item[]

1.- The {\em cosmic regime}, where
\bea
&{r\over r_\Lambda}\gg 1,\nonumber\\
&{r\over r_s}\gg 1.
\eea
The generic expansion should now read 
\bea
&A(r)=\frac{a_{-2}}{r^2}+\frac{a_{-3}}{r^3}+\frac{a_{-4}}{r^4}+\frac{a_{-5}}{r^5}+O\left(\frac{1}{r^6}\right),\nonumber\\
&B(r)=b_2 r^2+b_1 r+b_0+\frac{b_{-1}}{r}+O\left(\frac{1}{r^2}\right),
\eea 
and with the help of EoM we  obtain
\bea\label{sol22}
&A(r)=-\left(\frac{r_\Lambda}{r}\right)^2-\left(\frac{r_\Lambda}{r}\right)^4+\frac{a_{-5}}{r^5}+O\left(\frac{1}{r^6}\right),\nonumber\\
&\frac{B(r)}{b_2}= r^2-r_\Lambda^2+\frac{  a_{-5}}{r_\Lambda^2 r}+O\left(\frac{1}{r^2}\right).
\eea 
With the explicit form of these solutions and following the notation of the families near the origin, we call it the $(-2,2)$ family. It can be seen that this family of solutions contains one independent parameter $a_{-5}$ plus the cosmological constant (not counting  the trivial time rescaling $b_2$). Finally the Schwarzschild-de Sitter's  solution corresponds to the particular case
\be a_{-5}=r_s r_\Lambda^4, \quad \quad b_2=-\frac{1}{r_\Lambda^2}.\ee 

\item[]
2.- The {\em intermediate regime}, with
\bea
& {r\over r_s}\gg 1,\nonumber\\
&{r\over r_\Lambda} \ll 1.
\eea
Given the observational fact that when $r\sim 1 A.U.$,
\be
{r_s\over r}\sim 10^{-8}\equiv \e,
\ee
whereas
\be
{r\over r_\Lambda}\sim 10^{-15}\sim \e^2,
\ee
this is clearly the appropiate regime to study Solar system effects. 

%The adequate expension is
%\bea
%&B(r)=1-\e-\eta^2\nonumber\\
%&A(r)=1+\e+\e^2+\eta^2+\ldots
%\eea
We shall try then with the expansion
\bea
&B(r)=\sum_{n=0}^\infty {b_{-n}\over r^n},\nonumber\\
&A(r)=\sum_{n=0}^\infty {a_{-n}\over r^n}.
\eea
The result of such an asymptotic expansion is quite universal (that is, independent of the parameters $\a,\b,\g$ in \eqref{action2}), given by
\bea \frac{B(r)}{b_0}&=1+\frac{b_{-1}}{r},\nonumber\\
A(r)&=1-\frac{b_{-1}}{r}+\frac{b_{-1}^2}{r^2}-\frac{b_{-1}^3}{r^3}.\eea
This fact implies important consequences, which will be discussed in the conclusions. The particular case of Schwarzschild's  solution corresponds to  $b_{-1}=-r_s$.

\ei

\section{Expansion around an arbitrary point, $r_0$, in the second order approach}
%%%%%%%%%%%%%%%%%%%%%%%%%%%%%%%%%%%%%%%%%%%%%%%%%%%%%%%%%%%%%%%%%%%%%%%%%%%%%%%%%%%%%%%%%%%%%%%%%%%%%%%%%%%%%%%%
Now that we know the behavior of the functions near the origin and at infinity, we want to study whether they can be smoothly matched.  The only thing we need to check is that the solution in the neighborhood of an arbitrary point, $r_0$, has enough parameters to match it with the asymptotic expansion of either the $(2,2)$ or else the $(1,-1)$ solutions. We shall soon find that this is indeed the case.
\par
We assume that around the arbitrary point $r_0$, which has nothing to do with the radius of a possible horizon, the metric functions $A(r)$ and $B(r)$ are analytical and thus admit the following expansion
%
%
%All the observational evidence \cite{Will} (including binary pulsars?) supporting Schwarzschild's solution is also valid for the $(2,2)$ one. Very refined tests would be necessary to discriminate between the two.
\bea
& B(r)=b_0+b_1 (r-r_0)+b_2(r-r_0)^2+b_3(r-r_0)^3+b_4(r-r_0)^4+ O((r-r_0)^5) \nonumber \\
&A(r)=a_0+a_1 (r-r_0)+a_2(r-r_0)^2+a_3(r-r_0)^3 + O((r-r_0)^4).
\eea 
Given that the EoM \eqref{EOMC} and \eqref{EOMD} contain third derivatives of $A(r)$ and fourth derivatives of $B(r)$, we need to expand the functions up to that order to solve for the lowest order. After that, an iterative process can be used in order to get the higher-order terms of the solution. 

The lowest order of $H_{00}=0$ yields
\bea &-56a_1^3b_0^3r_0^3\left[2(\a+6\b)b_0-(\a-3\b)b_1r_0\right]-24a_0^4b_0^4(12\b+\g r_0^2)-\nonumber\\
&-4a_0^5b_0^4\left[4(\a-3\b)-6\g r_0^2+3\Lambda r_0^4\right]+a_0a_1b_0^2r_0^2\Big[104a_2b_0r_0\left[2(\a+6\b)b_0-(\a-3\b)b_1r_0\right]+\nonumber\\
&+a_1\left[4(7\a+60\b)b_0^2+57(\a-3\b)b_1^2r_0^2-4b_0r_0\left((13\a-84\b)b_1+38(\a-3\b)b_2r_0\right)\right]\Big]+\nonumber\\
&+a_0^3\Big[49 (\a - 3 \b) 
b_1^4r_0^4- 4 b_0b_1^2r_0^3 
((11 \a - 
78\b) b_1 + 
58 (\a - 3 \b) b_2r_0)-\nonumber\\
&-384(\a-3\b)b_0^3r_0^3(b_3+b_4r_0)+8b_0^4(2\a+30\b+3\g a_1r_0^3)-\nonumber\\
&-4b_0^2r_0^2\left[(5\a+12\b)b_1^2-36(\a-3\b)b_2^2r_0^2-4b_1r_0((13\a-66\b)b_2+18(\a-3\b)b_3r_0)\right]\Big]-\nonumber\\
&-2a_0^2b_0r_0\Big[8b_0r_0\left[3a_3b_0r_0(2(\a+6\b)b_0-(\a-3\b)b_1r_0)+\right.\nonumber\\
&\left.+a_2\left(2(\a+6\b)b_0^2+3(\a-3\b)b_1^2r_0^2+b_0r_0(-3(\a-6\b)b_1-8(\a-3\b)b_2r_0)\right)\right]-\nonumber\\
&-a_1\left[16(\a+6\b)b_0^3+29(\a-3\b)b_1^3r_0^3-12b_0b_1r_0^2((2\a-15\b)b_1+9(\a-3\b)b_2r_0)\right.+\nonumber\\
&\left.+4b_0^2r_0((-5\a+6\b)b_1+2r_0((13\a-48\b)b_2+18(\a-3\b)b_3r_0))\right]\Big]=0,\eea
and for $H_{11}=0$ we have
\bea
&7a_1^2b_0^2r_0^2\Big[4(\a-12\b)b_0^2-4(\a+6\b)b_0b_1r_0+(\a-3\b)b_1^2r_0^2\Big]+\nonumber\\
&+4a_0^4b_0^4\Big[4(\a-3\b)-6\g r_0^2+3\Lambda r_0^4\Big]+24a_0^3b_0^3\Big[\g b_1r_0^3+b_0(-12\b+\g r_0^2)\Big]-\nonumber\\
&-2a_0b_0r_0^2\Big[4a_2b_0\left[4(\a-12\b)b_0^2-4(\a+6\b)b_0b_1r_0+(\a-3\b)b_1^2r_0^2\right]-\nonumber\\
&-a_1\left[3(\a-3\b)b_1^3r_0^2-4b_0b_1r_0\left((2\a+3\b)b_1+2(\a-3\b)b_2r_0\right)+\right.\nonumber\\
&\left.+4b_0^2\left((\a+24\b)b_1+4(\a+6\b)b_2r_0\right)\right]\Big]-a_0^2\Big[16(\a-21\b)b_0^4-7(\a-3\b)b_1^4r_0^4+\nonumber\\
&+4b_0b_1^2r_0^3\left[(5\a+12\b)b_1+6(\a-3\b)b_2r_0\right]+32(\a+6\b)b_0^3r_0\left[-b_1+r_0(2b_2+3b_3r_0)\right]+\nonumber\\
&+4b_0^2r_0^2\left[(\a-48\b)b_1^2+4(\a-3\b)b_2^2r_0^2-12b_1r_0\left[2(\a+3\b)b_2+(\a-3\b)b_3r_0\right]\right]\Big]=0.
\eea
We can easily find a solution for this system of equations which is given by
\bea \label{b3}
&b_3=\frac{1}{48a_0^2b_0^2r_0^3\left[2(\a+6\b)b_0-(\a-3\b)b_1r_0\right]}\Big\{7a_1^2b_0^2r_0^2\Big[4(\a-12\b)b_0^2-\nonumber\\
&-4(\a+6\b)b_0b_1r_0+(\a-3\b)b_1^2r_0^2\Big]+4a_0^4b_0^4\left(4(\a-3\b)-6\g r_0^2+3\Lambda r_0^4\right)+\nonumber\\
&+24a_0^3b_0^3\left(\g b_1 r_0^3+b_0(-12\b+\g r_0^2)\right)+\nonumber\\
&+a_0^2\Big[-16(\a-21\b)b_0^4+7(\a-3\b)b_1^4 r_0^4+32(\a+6\b)b_0^3 r_0(b_1-2b_2 r_0)-\nonumber\\
&-4b_0 b_1^2 r_0^3\left((5\a+12\b)b_1+6(\a-3\b)b_2 r_0\right)-\nonumber\\
&-4b_0^2r_0^2\left((\a-48\b)b_1^2-24(\a+3\b)b_1b_2r_0+4(\a-3\b)b_2^3r_0^2\right)\Big]-\nonumber\\
&-2a_0b_0r_0^2\Big[4a_2b_0\left(4(\a-12\b)b_0^2-4(\a+6\b)b_0b_1r_0+(\a-3\b)b_1^2r_0^2\right)-\nonumber\\
&-a_1\left(3(\a-3\b)b_1^3r_0^2-4b_0b_1r_0((2\a+3\b)b_1+2(\a-3\b)b_2r_0)+\right.\nonumber\\
&\left.+4b_0^2((\a+24\b)b_1+4(\a+6\b)b_2r_0\right)\Big]\Big\},
\eea
and
\bea\label{b4}
&b_4=\frac{1}{384(\a-3\b)a_0^3b_0^3r_0^4\left[2(\a+6\b)b_0-(\a-3\b)b_1r_0\right]}\Big\{\nonumber\\
&-14a_1^3b_0^3r_0^3\left[4(\a^2+93\a\b+36\b^2)b_0^2-4(\a^2+3\a\b-18\b^2)b_0b_1r_0+(\a-3\b)^2b_1^2r_0^2\right]-\nonumber\\
&-4a_0^5b_0^4\left[2(5\a-6\b)b_0-7(\a-3\b)b_1r_0\right](4(\a-3\b)-6\g r_0^2+3\Lambda r_0^4)+\nonumber\\
&+a_0a_1b_0^2r_0^2\Big[8a_2 b_0r_0\Big(4 (7 \a^2 + 246 \a\b+ 252 \b^2) 
b_0^2- 
28 (\a^2 + 3\a\b - 18 \b^2) b_0b_1r_0 +\nonumber\\
&+ 7 (\a - 3 \b)^2 
b_1^2r_0^2
\Big) + 
a_1 \Big(-24 (7\a^2 - 174 \a\b + 216 \b^2) 
b_0^3 + 
21 (\a - 3 \b)^2 
b_1^3
r_0^3 - \nonumber\\
&-
2 (\a- 3 \b) b_0b_1r_0^2
((77 \a + 
318 \b) b_1- 
28 (\a - 3 \b) b_2r_0) + 4 
b_0^2r_0  ((77 \a^2 - 237 \a\b + 
828 \b^2) b_1 -\nonumber\\
&- 
28 (\a^2 + 3 \a \b - 18 \b^2) b_2r_0)\Big)\Big]+\nonumber\\
&+a_0^3\Big[-7(\a-3\b)^2b_1^5r_0^5-2(\a-3\b)b_0b_1^3r_0^4((17\a-78\b)b_1-44(\a-3\b)b_2r_0)+\nonumber\\
&+4b_0^2b_1r_0^3\Big((17\a^2+303\a\b-252\b^2)b_1^2+24(\a^2-9\a\b+18\b^2)b_1b_2r_0-60(\a-3\b)^2b_2^2r_0^2\Big)+\nonumber\\
&+8b_0^3r_0^2\Big((23\a^2-174\a\b+72\b^2)b_1^2-4(23\a^2+24\a\b-36\b^2)b_1b_2r_0+4(13\a^2+3\a\b-126\b^2)b_2^2r_0^2\Big)+\nonumber\\
&+16b_0^5\Big(2(5\a^2-75\a\b+342\b^2)-3a_1r_0(36(\a-3\b)\b+(-4\a\g+3\b\g)r_0^2)\Big)+\nonumber\\
&+8(\a-3\b)b_0^4r_0\Big(64(\a+6\b)b_2r_0+b_1(-46\a+30\b+15\g a_1r_0^3)\Big)\Big]+\nonumber\\
&+24a_0^4b_0^3\Big[6(\a-3\b)\g b_1^2r_0^4-(\a-3\b)b_0b_1r_0(60\b+\g r_0^2)+\nonumber\\
&+b_0^2\left[2(36(\a-6\b)\b+(-5\a\g+6\b\g)r_0^2)+(\a-3\b)a_1r_0(4(\a-3\b)-6\g r_0^2+3\Lambda r_0^4)\right]\Big]-\nonumber\\
&-4a_0^2b_0r_0\Big[a_1\Big(8(\a^2-96\a\b+117\b^2)b_0^4-5(\a-3\b)^2b_1^4r_0^4+\nonumber\\
&+(\a-3\b)b_0b_1^2r_0^3((25\a-12\b)b_1+6(\a-3\b)b_2r_0)+\nonumber\\
&+4b_0^3r_0^2(-(\a^2-78\a\b+144\b^2)b_1+2(7\a^2+30\a\b-72\b^2)b_2r_0)+\nonumber\\
&+4b_0^2r_0^2(-3(2\a^2+42\a\b-63\b^2)b_1^2+2(-8\a^2+3\a\b+63\b^2)b_1b_2r_0+6(\a-3\b)^2b_2^2r_0^2)\Big)+\nonumber\\
&+4b_0^2r_0\Big(3a_3r_0(-2(\a+6\b)b_0+(\a-3\b)b_1r_0)^2+a_2\left[-12(\a^2-24\a\b+36\b^2)b_0^2-\right.\nonumber\\
&\left.-(\a-3\b)b_1r_0^2(7(\a+6\b)b_1-8(\a-3\b)b_2r_0)+2b_0r_0((10\a^2-69\a\b+198\b^2)b_1-\right.\nonumber\\
&\left.-8(\a^2+3\a\b-18\b^2)b_2r_0)\right]\Big)\Big]\Big\}.
\eea

A remarkable property is that if we push the solution one order further in the expansion, what we find is that the solution depends on $a_0,a_1,a_2,a_4$ and $b_0,b_1,b_2$. We can see that we always get the first three parameters of each expansion $(a_0,a_1,a_2,b_0,b_1,b_2)$ plus the coefficient of the highest order of $A(r)$ (in this case $a_4$). This can be seen in the following way. If we take the system of equations for the lowest order they can be symbolically written as \bea 
&g(a_0,a_1,a_2,b_0,b_1,b_2, b_3)=0,\nonumber\\
&f(a_0,a_1,a_2,a_3,b_0,b_1,b_2,b_3,b_4)=0.
\eea
Therefore, we can always write the highest order of $B(r)$ in terms of the rest of the parameters using the second equation, and the second highest order of $B(r)$ in terms of the same ones using the first equation. As can be seen in equations \eqref{b3} and \eqref{b4}, $b_3$ and $b_4$ can be written in terms of the other seven parameters $(a_0,a_1,a_2,a_3)$ and $(b_0,b_1,b_2)$ plus the cosmological constant. Going to higher-order, it can be easily checked that we still have 7 independent parameters. In fact, only 6 of them are relevant as we can always reabsorb $b_0$ in a time rescaling. 
\par

We conclude that the general solution around an arbitrary point $r_0 \, (\gg r_s)$ depends on 6 arbitrary parameters plus the cosmological constant. This means that it can be smoothly matched (we have enough independent parameters) with the asymptotic $(-2,2)$ solution at ${r\over r_s} \rightarrow \infty$ as well as with {\em either} the $(2,2)$ solution or else the $(1,-1)$ solution at ${r\over r_s}\ll 1$. It also means that both the horizonless $(2,2)$ solution as well as the usual Schwarzschild-dS-like $(1,-1)$ solution match equally well with the asymptotic universal behavior.

%%%%%%%%%%%%%%%%%%%%%%%%%%%%%%%%%%%%%%%%%%%%%%%%%%%%%%%%%%%%%%%%%%%%%%%%%%%%%%%%%%%%%%%%%%%%%%%%%%%%%%%%%%%%%%%%%%%%%%%%%%%%%%%
\section{Generalities of the  first order approach}
%%%%%%%%%%%%%%%%%%%%%%%%%%%%%%%%%%%%%%%%%%%%%%%%%%%%%%%%%%%%%%%%%%%%%%%%%%%%%%%%%%%%%%%%%%%%%%%%%%%%%%%%%%%%%%%%
In the FO  approach, both the connection and the metric are {\em independent} fields. For Lagrangians linear in curvature (that is, the Einstein-Hilbert one) this viewpoint uncovers nothing new \cite{Anero}. Lagrangians quadratic (or higher) in curvature, are however essentially different in that FO and SO formalisms are not equivalent \cite{Olmo}. When the background metric is Ricci flat, for example, Minkowski or Schwarzschild spacetimes, the metric field obeys an algebraic (essentially trivial) EoM so that all the dynamics in encoded in the connection field. The spin content of the said connection field has been worked out in detail in \cite{AAGM}. Namely, one spin 3, three spin 2, five spin 1, and three spin 0. In those works, a systematic procedure for extracting the dominant graviton dynamics out of the connection field has been outlined.

We have already mentioned that quadratic theories imply ghosts, which manifest themselves classically as runaway solutions, among other inconsistencies. It has been conjectured that this is not necessarily true in the FO approach. There the graviton field in a flat background obeys an algebraic EoM, (and in particular, there is no kinetic energy term for the graviton) and the whole dynamics is encoded in the connection field, which is akin to a  three-index gauge field. Indeed, both the presence of spin 3 and several spin 2, point towards inconsistencies in the quantum theory, but more work is needed to clarify this issue.
\par
The general Lagrangian in the FO approach has many terms, which have been worked out in detail in \cite{AAGM}. This is due to the fact that when there is {\em nonmetricity} (that is whenever the connection is not Levi-Civita) the Riemann tensor has fewer symmetries than in the standard case so that there is correspondingly a greater number of independent quadratic operators in the Lagrangian. The general analysis becomes somewhat laborious. In this work, however, we content ourselves to study a simple Lagrangian, as close as possible to the one we analyzed in the SO approach.  In the same vein, we will not look for nonmetric solutions of the equations; but rather {\em assume} that the background connection is the Levi-Civita one, and study nonmetric connections only as perturbations when this is necessary.
\par
To understand the restricted dynamics of the connection field, that is, to explore  some general aspects of the huge space of classical solutions in FO,  we return to the old Lagrangian
\be\label{action3}
S=\int d^4 x \sqrt{|g|} \, \Big[-\Lambda-\g \, R-2 \a \, R_{\m\n}^2+\left(\b+{2\over 3} \a \right)\,R^2\Big],
\ee
and analyze it in the restricted FO formalism. We assume that the metric and the connection are independent fields, but {\em after} getting the EoM, we fix the background connection to the Levi-Civita one. This is reminiscent of the usual 1.5 rule extensively employed in supergravity \cite{PVN}.
\par
Taking the following background expansion (where the presence of $\kappa$ assigns the correct engineering dimensions to the graviton fluctuations)
\bea
g_{\m\n}&=\bg_{\m\n}+\kappa h_{\m\n}\nonumber\\
\Gamma^{\lambda}_{\m\n}&=\widebar{\Gamma}^{\lambda}_{\m\n}+A^{\lambda}_{\m\n},
\eea	
the EoM for the graviton reads
\bea\label{EOMFOG} &H^{FO}_{\m\n}=\g\left(\bR_{\m\n}-\frac{1}{2}\bg_{\m\n}\bR\right)-2\left(\b+\frac{2\a}{3}\right)\bR\left(\bR_{\m\n}-\frac{1}{4}\bg_{\m\n}\bR\right)+4\a\left(\widebar{R}_{\m\l}\widebar{R}_\n^{~\l}-\frac{1}{4}\bg_{\m\n}\bR_{\r\s}^2\right)-\frac{1}{2}\bg_{\m\n}\Lambda=0,\eea
and the EoM for the connection takes the form
\bea\label{EOMFOC}
&\Big\{-\g\Big[\widebar{g}^{\m\n}\widebar{\nabla}_\lambda-\frac{1}{2}\left(\delta^{\n}_{\lambda}\widebar{\nabla}^\m+\delta^{\m}_{\lambda}\widebar{\nabla}^\n\right)\Big]-4\a\Big[\widebar{R}^{\m\n}\widebar{\nabla}_\lambda-\frac{1}{2}\left(\widebar{R}^{\m\t}\delta^{\n}_{\lambda}\widebar{\nabla}_\t+\widebar{R}^{\n\t}\delta^{\m}_{\lambda}\widebar{\nabla}_\t\right)\Big]+\nonumber\\
&+2\left(\b+{2\over 3} \a \right)\widebar{R}\Big[\widebar{g}^{\m\n}\widebar{\nabla}_\lambda-\frac{1}{2}\left(\delta^{\n}_{\lambda}\widebar{\nabla}^\m+\delta^{\m}_{\lambda}\widebar{\nabla}^\n\right)\Big]\Big\}A^{\lambda}_{\m\n}=0.\eea
We can integrate by parts in the equation for the connection and use the metricity condition, $\widebar{\nabla}_\lambda\bg_{\m\n}=0$,  together with the Bianchi identity so that we obtain the following system of equations
\bea\label{FOEOM} &H^{FO}_{\m\n}=\g\left(\bR_{\m\n}-\frac{1}{2}\bg_{\m\n}\bR\right)-2\left(\b+\frac{2\a}{3}\right)\bR\left(\bR_{\m\n}-\frac{1}{4}\bg_{\m\n}\bR\right)+4\a\left(\widebar{R}_{\m\l}\widebar{R}_\n^{~\l}-\frac{1}{4}\bg_{\m\n}\bR_{\r\s}^2\right)-\frac{1}{2}\bg_{\m\n}\Lambda=0,\nonumber\\
&C_{\l}^{\m\n}=4\a\Big[\widebar{\nabla}_\lambda\widebar{R}^{\m\n}-\frac{1}{4}\left(\delta^{\n}_{\lambda}\widebar{\nabla}^\m\widebar{R}+\delta^{\m}_{\lambda}\widebar{\nabla}^\n\widebar{R}\right)\Big]-2\left(\b+{2\over 3} \a \right)\Big[\widebar{g}^{\m\n}\widebar{\nabla}_\lambda\widebar{R}-\frac{1}{2}\left(\delta^{\n}_{\lambda}\widebar{\nabla}^\m\widebar{R}+\delta^{\m}_{\lambda}\widebar{\nabla}^\n\widebar{R}\right)\Big]=0.
\eea
Let us note that Minkowski, Schwarzschild-de Sitter and in fact any constant curvature space with $R=-\frac{\Lambda}{2\g}$ are solutions of these equations.

One can study these two equations order by order in $\kappa$. We will see however that in this restricted FO approach, the lowest order of the EoM is tautological for several simple spacetimes, so that we will need to study perturbatively nonmetric solutions, solving the EoM for the perturbation of the connection field $A^{\lambda}_{\m\n}$.  We can then extract the metric perturbation from the connection perturbation using a procedure detailed in the following section.

%%%%%%%%%%%%%%%%%%%%%%%%%%%%%%%%%%%%%%%%%%%%%%%%%%%%%%%%%%%%%%%%%%%%%%%%%%%%%%%%%%%%%%%%%%%%%%%%%%%%%%%%%%%%%%%%%%%%%%%%%%%%%%
\subsection{How to derive a spacetime metric from the connection field}
%%%%%%%%%%%%%%%%%%%%%%%%%%%%%%%%%%%%%%%%%%%%%%%%%%%%%%%%%%%%%%%%%%%%%%%%%%%%%%%%%%%%%%%%%%%%%%%%%%%%%%%%%%%%%%%%%%%%%%%%%%%%%%%
It has been pointed out that in theories treated in the FO formalism all physics is sometimes (for example, whenever the background spacetime is Ricci flat) conveyed by the connection field, so we have to be able to recover the physical space-time metric from the said connection field. Nevertheless, not every connection is metric compatible; that is, it is not always possible to find a metric such that the given connection (even assumed to be torsion-free) is the Levi-Civita one stemming from the metric itself. 
\par
The condition for that to be true can be  clearly stated using the metricity condition for a general connection, $\nabla_\lambda g_{\delta \beta} =0$, which gives
%Christoffel's symbols of
%first kind, namely 
%\be 
%\pd_\m\bigg(\left\{\d;\b\l\right\}+\left\{\b;\l \d\right\}\bigg)=\pd_\l\bigg(\left\{\d;\b
%\m \right\}+\left\{\b;\d \m\right\}\bigg) 
%\ee
% which expresses the obvious fact that 
%\be 
%\pd_\m\pd_\l~ g_{\d\b}=\pd_\l\pd_\m
%~g_{\b\d} 
%\ee 
%but when we express this condition is in terms of second kind symbols it involves not only the connection, but also the metric tensor:
%\be
%\pd_\m\left(g_{\d\r} \Gamma^\r_{\b\l}+g_{\b\r} \Gamma^\r_{\l\d}\right)=\pd_\l\left(g_{\d\r} \Gamma^\r_{\b\m}+g_{\b\r} \Gamma^\r_{\m\d}\right)
%\ee
%In order to determine the generated metric in such
%cases as it exists, (that is, when the integrability condition is
%fulfilled), 
the linear  system of partial differential equations 
\be 
\pd_\l~g_{\d\b}=g_{\a\d}~ \Gamma^\a_{\b\l}+ g_{\a\b}~\Gamma^\a_{\l\d}.
\ee
%whose trace implies
%\be
%g^{\d\b}~\pd_\l~ g_{\d\b}= 2 ~\Gamma^\b_{\b\l}
%\ee
The key step in this construction is to use this compatibility condition even for connections that are not metric compatible. This is a particular way of choosing a projection of the three-index connection onto a spin two part. 
This is in general a quite complicated system for the ten unknown functions $g_{\a\b}$ in terms of the $\Gamma^\m_{\a\b}$ but we can solve it order by order in $\kappa$ expanding the metric and the connection
\bea
&g_{\a\b}=\bg_{\a\b}+\kappa h_{\a\b},\nonumber\\
&\Gamma^\a_{\b\g}=\overline{\Gamma}^\a_{\b\g}+A^\a_{\b\g},
\eea
where we are assuming that $A^\a_{\b\g}=O(\kappa)$ and we take the Levi-Civita connection as the background of the connection field (namely, we work in the restricted FO formalism). The preceding equation up to order $\kappa$ then reads
\bea
&\pd_\l\bg_{\d\b}+\kappa\pd_\l~h_{\d\b}= \bg_{\a\d} \overline{\Gamma}^\a_{\b\l}+\bg_{\a\b} \overline{\Gamma}^\a_{\l\d}+\kappa h_{\a\d}\overline{\Gamma}^\a_{\b\l}+\kappa h_{\a\b} \overline{\Gamma}^\a_{\l\d}+\bg_{\a\d} A^\a_{\b\l}+\bg_{\a\b} A^\a_{\l\d}.
\eea
Using the metricity property of the Levi-Civita connection $\bn_\l\bg_{\d\b}=0$, we have
\be\label{CC}
\kappa \bn_\l h_{\a\b}=A_{\a|\b\l}+A_{\b|\a\l}.
\ee
Please note that $A_{\d|\b\l}=A_{\d|\l\b}$, but there is no symmetry in general affecting the first index.
Based on this, we shall {\em define} (even when this condition is not satisfied), the metric associated to a given connection by
\be\label{EMA}
\kappa \bar{\Box}h_{\a\b}=\bn^\l\left(A_{\a|\b\l}+A_{\b|\a\l}\right),
\ee
where we have just taken the derivative of the previous equation. This is nothing else than projecting a  spin-2 part from a general torsionless connection.
%%%%%%%%%%%%%%%%%%%%%%%%%%%%%%%%%%%%%%%%%%%%%%%%%%%%%%%%%%%%%%%%%%%%%%%%%%%%%%%%%%%%%%%%%%%%%%%%%%%%%%%%%%%%%%%%%%%%%%%%%%%%%%%
\subsection{Perturbations of constant curvature spacetimes (CCS)}
%%%%%%%%%%%%%%%%%%%%%%%%%%%%%%%%%%%%%%%%%%%%%%%%%%%%%%%%%%%%%%%%%%%%%%%%%%%%%%%%%%%%%%%%%%%%%%%%%%%%%%%%%%%%%%%%
In order to explore the procedure analyzed in the previous section, let us begin by exploring a toy model, namely, a maximally symmetric spacetime. These spacetimes are defined as the ones for which their Riemann tensor has the simple form given by 
\be 
\bR_{\a\b\g\d}=-{ L\over
	3}\left(\bg_{\a\g}\bg_{\b\d}-\bg_{\a\d}\bg_{\b\g}\right).
\label{RCCS}
\ee
In four dimensions, the EoM \eqref{EOMFOG} for the metric just yields the well-known relationship between the curvature and the cosmological constant given by
\be L=\frac{\Lambda}{2\g},\ee
so that the graviton EoM does not give us more information. In the restricted FO approach the EoM for the connection \eqref{EOMFOC} takes the form
\be \label{AA}\Big[\widebar{g}^{\m\n}\widebar{\nabla}_\lambda-\frac{1}{2}\left(\delta^{\n}_{\lambda}\widebar{\nabla}^\m+\delta^{\m}_{\lambda}\widebar{\nabla}^\n\right)\Big]A^{\lambda}_{\m\n}=0.
\ee
The aim of this section is to analyze perturbatively solutions for more general connections, in particular, connections with nonmetricity. Given the tautological character of the graviton EoM \eqref{EOMFOG} we can then extract the spin 2 part from the connection perturbation. 
\par
In the particular case of maximally symmetric spacetimes, it will prove convenient  to work with Synge's \cite{Synge} world function $\s(x,y)$ (see Appendix \ref{C} to see its definition and the properties that follow). In the Riemannian case, it is equivalent and sometimes simpler to use the geodesic distance, $s(x,\xp)$.
\par
In the following, we shall discuss several ansatzes for the connection, which we shall classify according to the behavior under the contraction with $s_\lambda$. For each sector, we also obtain the form of the metric perturbations stemming from the connection perturbation using the relation \eqref{EMA}. It adds no difficulty to perform most calculations for arbitrary elliptic spacetime dimension, $n$.
%%%%%%%%%%%%%%%%%%%%%%%%%%%%%%%%%%%%%%%%%%%%%%%%%%%%%%%%%%%%%%%%%%%%%%%%%%%%%%%%%%%%%%%%%%%%%%%%%%%%%%%%%%%%%%%%%%%%%%%%%%%%%%
\subsubsection{Longitudinal connections}
%%%%%%%%%%%%%%%%%%%%%%%%%%%%%%%%%%%%%%%%%%%%%%%%%%%%%%%%%%%%%%%%%%%%%%%%%%%%%%%%%%%%%%
We start with the ansatz $A^{\lambda}_{\m\n}=f(s)s^\l s_\m s_\n$, which can be checked to be a particular solution of \eqref{AA} for any $f(s)$\footnote{Let us note that this ansatz would correspond to a scalar sector in a rest frame defined by $s^\m=\d^\m_0$, so that this solution reduces to $A_{000}$ and represents a spin $0$ scalar mode. We are not getting into more detail here as the purpose of the paper is not the analysis of the spin components of the connection field. Nevertheless, a detailed derivation of the different spin projectors and the combination of the components of the connection that correspond to each of the spin components can be found in \cite{AlvarezAneroSantos}.}. Here, $s_\m$ stands for the derivatives of the arc distance $s$, and the indices are raised with the arbitrary CCS metric appearing in \eqref{RCCS}. The properties of these derivatives can be found in Appendix \ref{C}, where one can check that they can be written in terms of a function of the arc distance itself and the constant curvature scale $L$.  
\par
Introducing the ansatz in the rhs of the compatibility condition \eqref{EMA} we get
\be
\bn^\l\left(A_{\m|\n\l}+A_{\n|\m\l}\right)=2\left(f^\prime(s)+f(s)\bar{\Box}s\right)s_\m s_\n,
\ee
where prime denotes derivation with respect to $s$. Let us assume that the perturbation of the metric is also a function of the geometric quantities only, that is, a function of the arc length. This is obviously not true for any possible perturbation, but we stick to this choice in order to get to simpler results. This is equivalent to taking the following ansatz for the metric perturbation
\be
h_{\m\n}=h_0(s) \bg_{\m\n}+h_1(s) s_\m s_\n,
\ee
so that the lhs  of \eqref{EMA} reads (for arbitrary spacetime dimension, $n$) 
\bea
\bar{\Box} h_{\m\n}&=\left[h_0^{''}(s)+h_0^{'}(s){n-1\over L\tan\,{s\over L}}\right]\bg_{\m\n}+\left[h_1^{''}(s)+h_1^{'}(s){n-1\over L\tan\,{s\over L}}\right]s_{\m}s_{\n}+\nonumber\\
&+h_1(s)\Big[{2\over L^2\tan^2\,{s\over L}}\left(\bg_{\m\n}-s_\m s_\n\right)-2 {n-1\over L^2\tan^2\,{s\over L}} s_\m s_\n\Big].
\eea
With this, the compatibility condition \eqref{EMA} establishes the system of ODE
\bea
&h_0^{''}(s)+h_0^{'}(s){n-1\over L\tan\,{s\over L}}+{2\over L^2\tan^2\,{s\over L}}h_1(s)=0\nonumber\\
&h_1^{''}(s)+h_1^{'}(s){n-1\over L\tan\,{s\over L}}-{2n\over L^2\tan^2\,{s\over L}}h_1(s)=\frac{2}{\kappa}\Big[f^\prime(s)+{n-1\over L\tan\,{s\over L}} f(s)\Big].
\label{ODE}
\eea
We can then obtain a solution to the metric perturbations as a function of the arc length. Let us remember that the function $f(s)$ appearing in the ansatz of the connection is arbitrary at this point, but we can choose a certain form of this function in order to get a simple solution for these equations.
%%%%%%%%%%%%%%%%%%%%%%%%%%%%%%%%%%%%%%%%%%%%%%%%%%%%%%%%%%%%%%%%%%%%%%%%%%%%%%%%%%%%%%%%%%%%%%%%%%%%%%%%%%%%%%%%%%%%%%%%%%%%%%%
%\subsubsection{Solution near the origin}
%%%%%%%%%%%%%%%%%%%%%%%%%%%%%%%%%%%%%%%%%%%%%%%%%%%%%%%%%%%%%%%%%%%%%%%%%%%%%%%%%%%%%%%%%%%%%%%%%%%%%%%%%%%%%%%%%%%%%%%%%%%%%%
\par
Let us begin by considering $s\ll L$ so that the ODE  reduce to
\bea
&h_0^{''}(s)+h_0^{'}(s){n-1\over s}+{2\over s^2}\,h_1(s)=0\nonumber\\
&h_1^{''}(s)+h_1^{'}(s){n-1\over s}-{2n\over s^2 }\,h_1(s)=\frac{2}{\kappa}\Big[f^\prime(s)+{n-1\over s} f(s)\Big].
\eea
Assuming a particular sector where the arbitrary function takes the precise form
\be
f(s)=-{C\over s},
\ee
we can find a solution given by
\bea
h_1(s)&=\frac{C}{\kappa n}(n-2)+C_1s^{-n}+C_2s^2 \nonumber \\
h_0(s)&=-\frac{C_1}{n}s^{-n}-\frac{C_2}{n}s^2-\frac{C_3}{n-2}s^{2-n}-\frac{2C}{n\kappa}\log s.\eea
%%%%%%%%%%%%%%%%%%%%%%%%%%%%%%%%%%%%%%%%%%%%%%%%%%%%%%%%%%%%%%%%%%%%%%%%%%%%%%%%%%%%%%%%%%%%%%%%%%%%%%%%%%%%%%%%%%%%%%%%%%%%%%%
%\subsubsection{An exact solution}
%%%%%%%%%%%%%%%%%%%%%%%%%%%%%%%%%%%%%%%%%%%%%%%%%%%%%%%%%%%%%%%%%%%%%%%%%%%%%%%%%%%%%%%%%%%%%%%%%%%%%%%%%%%%%%%%%%

We can also find an exact solution of the original equations performing the change
\be
x\equiv \cos\,\frac{s}{L}.
\ee
We can rewrite the homogeneous part of equations \eqref{ODE} as
\bea\label{H}
&(1-x^2){\pd^2 h_0\over \pd x^2}- n\,x {\pd h_0\over \pd x} +2 {x^2\over 1-x^2} h_1=0\nonumber\\
&(1-x^2){\pd^2 h_1\over \pd x^2}-  n\, x {\pd h_1\over \pd x} -2n {x^2\over 1-x^2} h_1=0,\nonumber\\
\eea
and the solution of the second of those is then given by
\be
h_1(x)= \left(x^2-1\right)^{\frac{2-n}{4}}\Big[ c_1 P_{\frac{n+2}{2}}^{\frac{-1+\sqrt{1+6n+n^2}}{2}}(x)+c_2  Q_{\frac{n+2}{2}}^{\frac{-1+\sqrt{1+6n+n^2}}{2}}(x)\Big].
\ee
On the other hand, we can combine the equations \eqref{H}
\be
(1-x^2){\pd^2 \over \pd x^2}\left(h_1+nh_0\right)- n x {\pd \over \pd x}\left(h_1+nh_0\right)=0
\ee
whose solution is
\be
h_1(x)+nh_0(x)= -c_1 x \left(x^2-1\right)^{(2-n)/2} \, _2F_1\left(1;\frac{3-n}{2};\frac{3}{2};x^2\right)+c_2.
\ee
It remains  to correct for the non-homogeneous pieces, but taking into account that $f(s)$ was arbitrary, we can choose
\be
f(s)= C \left(\sin\,{s\over L}\right)^{-(n-1)}.
\ee
In this way, the second member of the inhomogeneous equation vanishes, and the quoted previous solution carries on. Finally we get
\bea
&h_0(x)=\frac{1}{n}\Big[ -c_1 x \left(x^2-1\right)^{(2-n)/2} \, _2F_1\left(1;\frac{3-n}{2};\frac{3}{2};x^2\right)+c_2-h_1(x)\Big],\nonumber\\
&h_1(x)= \left(x^2-1\right)^{\frac{2-n}{4}}\Big[ c_1 P_{\frac{n+2}{2}}^{\frac{-1+\sqrt{1+6n+n^2}}{2}}(x)+c_2  Q_{\frac{n+2}{2}}^{\frac{-1+\sqrt{1+6n+n^2}}{2}}(x)\Big].\eea

%%%%%%%%%%%%%%%%%%%%%%%%%%%%%%%%%%%%%%%%%%%%%%%%%%%%%%%%%%%%%%%%%%%%%%%%%%%%%%%%%%%%%%%%%%%%%%%%%%%%%%%%%%%%%%%%%%%%%%%%%%%%%%%
\subsubsection{Transverse connection} 
%%%%%%%%%%%%%%%%%%%%%%%%%%%%%%%%%%%%%%%%%%%%%%%%%%%%%%%%%%%%%%%%%%%%%%%%%%%%%%%%%%%%%%%%%%%%%%%%%%%%%%%%%%%%%%%%%%%%%%%%%%%%
In the following, we further explore another possible ansatz for the connection field given by\footnote{Taking the rest frame as before this would correspond to a spin 2 mode that could be used to obtain the metric degree of freedom describing the graviton \cite{AlvarezAneroSantos}. But again, we shall not get into the details of the derivation of the precise form of the spin components.}
\be
A^{\lambda}_{\m\n}=g(s)s^\l s_{\m\n}.
\ee
We call these connections transverse because of the vanishing contraction $s^\m A^{\lambda}_{\m\n}=0$. 
%It can be checked that in the rest frame, commented above, this ansatz would collapse to a spin 2 component \cite{AlvarezAneroSantos}.
Introducing this ansatz in the EoM of the connection \eqref{AA}, we get 
\bea 
&g^{'}(s)s^2\bar{\Box} s+g(s)(\bar{\Box}s)^2+g(s)s^\l\bar{\Box} s_\l=0.\eea
We can use the properties of the arc length in \eqref{P} 
to rewrite the ODE for $g(s)$ as
\be g^{'}(s){n-1\over L~\tan~{s\over L}}+g(s)\left({n-1\over L~\tan~{s\over L}}\right)^2-g(s){n-1\over L^2~\sin^2~{s\over L}}=0.\ee
We can solve it for obtaining
\bea\label{gs} g(s)&=\frac{C_1}{2^{(n-1)/2}\cos(s/L)(\sin(s/L))^{(n-2)}}.\eea
We can see that in this case, the function appearing in the ansatz of the connection is constrained to have a particular form as a function of the arc length. 
\par
The next step is to substitute the compatibility condition \eqref{EMA}, 
\be
\bn^\l\left(A_{\m|\n\l}+A_{\n|\m\l}\right)=g(s)\Big[{2\over L^2\tan^2\,{s\over L}}\left(\bg_{\m\n}-s_\m s_\n\right)-2 {n-1\over L^2\tan^2\,{s\over L}} s_\m s_\n\Big],
\ee
yielding the system of ODE
\bea
&h_0^{''}(s)+h_0^{'}(s){n-1\over L\tan\,{s\over L}}+{2\over L^2\tan^2\,{s\over L}}h_1(s)=\frac{g(s)}{\kappa}{2\over L^2\tan^2\,{s\over L}},\nonumber\\
&h_1^{''}(s)+h_1^{'}(s){n-1\over L\tan\,{s\over L}}-{2n\over L^2\tan^2\,{s\over L}}h_1(s)=-\frac{g(s)}{\kappa}{2n\over L^2\tan^2\,{s\over L}}.
\eea
We already know the solution of the corresponding homogeneous equations, as the lhs of the equations remains unchanged, which was obtained for the scalar sector. Let us now introduce the non-homogeneous part.
%%%%%%%%%%%%%%%%%%%%%%%%%%%%%%%%%%%%%%%%%%%%%%%%%%%%%%%%%%%%%%%%%%%%%%%%%%%%%%%%%%%%%%%%%%%%%%%%%%%%%%%%%%%%%%%%%%%%%%%%%%%%%
%\subsubsection{Solution near the origin}
%%%%%%%%%%%%%%%%%%%%%%%%%%%%%%%%%%%%%%%%%%%%%%%%%%%%%%%%%%%%%%%%%%%%%%%%%%%%%%%%%%%%%%%%%%%%%%%%%%%%%%%%%%%%%%%%%%%%%%%%%%%
Again, one can consider the regime $s\ll L$ so that the particular function appearing in the ansatz can be Taylor expanded as
\be
g(s)={C\kappa\over s^{n-2}}.
\ee
We can finally solve for the metric functions and get
\bea
h_1(s)&=\frac{C}{s^{n-2}}+\frac{C_1}{s^n}+\frac{C_2}{s^2}, \nonumber \\
h_0(s)&=-\frac{C_1}{n}s^{-n}-\frac{C_2}{n}s^2-\frac{C_3}{n-2}s^{2-n}+C_4.
\eea
Let us mention again that these are very particular solutions to illustrate the procedure of obtaining nonmetric connections from where one could extract a spin 2 sector including the graviton solution.  In the following sections, we move towards more physically interesting solutions, focusing on black hole solutions and analyzing them in the FO formalism.
%%%%%%%%%%%%%%%%%%%%%%%%%%%%%%%%%%%%%%%%%%%%%%%%%%%%%%%%%%%%%%%%%%%%%%%%%%%%%%%%%%%%%%%%%%%%%%%%%%%%%%%%%%%%%%%%%%%%%%%%%%%%%%
\section{Structural stability of  Schwarzschild's  solution in the restricted first order approach}
%%%%%%%%%%%%%%%%%%%%%%%%%%%%%%%%%%%%%%%%%%%%%%%%%%%%%%%%%%%%%%%%%%%%%%%%%%%%%%%%%%%%%%%%%
Let us now analyze the stability of the Schwarzschild metric from our new vantage point.  It is convenient to do so, to change slightly the action principle by including explicitly a Riemann square term in the Lagrangian and quenching the cosmological constant to zero. We start then with the action given by
\be
S=\int d^4 x \sqrt{|g|} \, \Big[-\g \, R-2 \a \, R_{\m\n}^2+\left(\b+{2\over 3} \a \right)\,R^2+\l R_{\a\b\r\s}^2\Big].
\ee
For this action, the EoM reduce to
\bea &\left[\frac{1}{2}\bg_{\m\n}\bR_{\a\b\r\s}^2-2
\widebar{R}_{\m\l\rho\sigma}\widebar{R}_\n^{~\l\rho\sigma}\right]h^{\m\n}=0\nonumber\\
&\Big[\widebar{R}_{\lambda}^{~\m\t\n}\widebar{\nabla}_\t+\widebar{R}_{\lambda}^{~\n\t\m}\widebar{\nabla}_\t\Big]A^{\lambda}_{\m\n}=0	\eea
The order zero part of these equations, that is, the background EoM, is tautological when taking as the background fields the Levi-Civita  connection and the Schwarzschild metric. We can then focus on obtaining perturbatively order $\kappa$ solutions, which lead to a nonmetric connection.  One can easily realize that the EoM  for the metric is exactly the Bach-Lanczos identity \cite{Bach} for Ricci-flat metrics where Weyl's tensor is exactly the Riemann tensor (a simple proof of it was included as an appendix in \cite{AAGM})
\be\frac{1}{2}\bg_{\m\n}W_{\a\b\r\s}^2-2
W_{\m\l\rho\sigma}W_\n^{~\l\rho\sigma}=0.\ee
Thus, the order $\kappa$ piece of the graviton EoM is also tautological. We are here in the situation where the EoM for the perturbations of the metric is empty (just because it is a geometric identity). We shall then try to get some information on those perturbations through the  EoM for the connection field. Even if this is a somewhat ambiguous procedure given the high spin content of this field, physically, we are just selecting a spin 2 projection of the general connection field. 
\par
The EoM  for the connection in turn, reads
\bea\label{EoM} 
&\widebar{R}_{\lambda}^{~\m\t\n}\widebar{\nabla}_\t A^{\lambda}_{\m\n}=0	
\eea
We will focus on two particular ansatzes for the connection and metric perturbations and use the projection \eqref{CC} of the connection onto a spin 2 part given by the metric perturbations.
%%%%%%%%%%%%%%%%%%%%%%%%%%%%%%%%%%%%%%%%%%%%%%%%%%%%%%%%%%%%%%%%%%%%%%%%%%%%%%%%%%%%%%%%%%%%%%%%%%%%%%%%%%%%%%%%%%%%%%%%%%%%%%%
%\subsection{One temporal  perturbation of the metric}
%%%%%%%%%%%%%%%%%%%%%%%%%%%%%%%%%%%%%%%%%%%%%%%%%%%%%%%%%%%%%%%%%%%%%%%%%%%%%%%%%%%%%%%%%%%%%%%%%%%%%%%%%%%%%%%%%%%%%%%%%%%%%
\begin{itemize}
	\item First, we assume a perturbation of the spacetime metric $h_{\m\n}$ that only affects the time-time component of the perturbation and just depends on one arbitrary function 
	\be 
	h_{00}=B(r).
	\ee 
	To determine this perturbation, one can see that with the compatibility condition \eqref{CC} we only need three components of the connection field. Namely, $A^0_{01}=A^0_{01}$  and $A^1_{00}$.
	We now take an ansatz for the connection field inspired by the symmetry and form of the components of the Levi-Civita connection for a Schwarzschild metric. In this case, we take
	\be
	A^0_{01}=A^0_{01}=f(r),  \quad \quad A^1_{00}=g(r).
	\ee
	With this connection, the EoM \eqref{EoM} reads
	\bea\label{2} &\frac{r_s}{r^5(r-r_s)^2}\Big\{(r-2r_s)(r-r_s)^2f(r)+r\Big[r^2g(r)-(r-r_s)\left((r-r_s)^2f'(r)+r^2g'(r)\right)\Big]\Big\}=0.	\eea
	Finally, introducing this form of the connection in the compatibility condition \eqref{CC}, we obtain a system of equations given by
	\bea\label{cc2}
	&-\frac{\kappa r_s}{2(r-r_s)}B(r)=\frac{r-r_s}{r}f(r)-\frac{r}{r-r_s}g(r),\nonumber\\
	&\kappa\Big[B'(r)-\frac{ r_s}{(r-r_s)}B(r)\Big]=\frac{2(r-r_s)}{r}f(r).
	\eea
	Let us note that when $f(r)=g(r)$, the equation of motion and the compatibility condition form an incompatible system of equations. Combining both equations of \eqref{cc2} and using the EoM \eqref{2}, we can solve for the function appearing in the metric perturbation 
	\bea\label{B} B(r)&=C_1\sqrt{1-\frac{r_s}{r}}+\frac{C_2}{4}\sqrt{1-\frac{r_s}{r}}\Big[\left(2r+3r_s\right)\sqrt{r(r-r_s)}+3r_s^2\log\left(\sqrt{r}+\sqrt{r-r_s}\right)\Big].\eea
	Going back to the system of equations \eqref{cc2} the solution of the connection perturbations is given by
	\bea f(r)&=-\frac{\kappa}{16\sqrt{r}(r-r_s)^{3/2}}\Big[4C_1r_s+\sqrt{r(r-r_s)}\left(-8r^2+2r\,r_s+3r_s^2\right)C_2+3C_2r_s^3\log\left(\sqrt{r}+\sqrt{r-r_s}\right)\Big],\nonumber\\
	g(r)&=\frac{\kappa}{16r^{5/2}}\Big[4C_1r_s\sqrt{r-r_s}+\sqrt{r}\left(8r^3-6r^2\,r_s+r\,r_s^2-3r_s^3\right)C_2+3C_2r_s^3\sqrt{r-r_s}\log\left(\sqrt{r}+\sqrt{r-r_s}\right)\Big].\eea
	The only perturbation of the metric is then encoded in the $g_{00}$ component which reads 
	\bea
	g_{00}&=\left(1-\frac{r_s}{r}\right)+\kappa\Big\{C_1\sqrt{1-\frac{r_s}{r}}+\frac{C_2}{4}\sqrt{1-\frac{r_s}{r}}\Big[\left(2r+3r_s\right)\sqrt{r(r-r_s)}+3r_s^2\log\left(\sqrt{r}+\sqrt{r-r_s}\right)\Big]\Big\}
	\eea
	In conclusion, this is a general result of this type of perturbative analysis. We always recover a horizon, unperturbed at the initial location,  $r=r_s$.
	%%%%%%%%%%%%%%%%%%%%%%%%%%%%%%%%%%%%%%%%%%%%%%%%%%%%%%%%%%%%%%%%%%%%%%%%%%%%%%%%%%%%%%%%%%%%%%%%%%%%%%%%%%%%%%%%%%%%%%%%%%%%%%%
	%\subsection{A spacetime  perturbation}
	%%%%%%%%%%%%%%%%%%%%%%%%%%%%%%%%%%%%%%%%%%%%%%%%%%%%%%%%%%%%%%%%%%%%%%%%%%%%%%%%%%%%%%%%%%%%%%%%%%%%%%%%%%%%%%%%%%%%%%%%%%%%%%%
	\item Next, we assume a metric perturbation depending on two arbitrary functions
	\be
	h_{00}=B(r), \quad \quad h_{11}=-A(r).
	\ee
	In order to determine the spacetime metric perturbations, we only need six components of the connection. In this case, these are 
	\begin{align}
	A^0_{01},&=A^0_{01}=f(r),  &A^1_{00}&=g(r), &A^1_{11}&=c(r),&A^1_{22}&=e(r), &A^1_{33}&=e(r)\sin^2\theta.
	\end{align}
	Introducing these functions in the connection EoM \eqref{EoM}, we have
	\bea\label{3} &\frac{r_s}{r^6(r-r_s)^2}\Big\{r(r-2r_s)(r-r_s)^2f(r)+r^4g(r)-\nonumber\\
	&-(r-r_s)\Big[-3(r-r_s)e(r)+r\left(r(r-r_s)^2f'(r)+r^3g'(r)+(r-r_s)e'(r)\right)\Big]\Big\}=0.	\eea
	Besides, the compatibility condition \eqref{CC} reduces to the following system of equations
	\bea\label{cc3}&\kappa\Big[\frac{(r-r_s)r_s}{2r^3}A(r)-\frac{r_s}{2r(r-r_s)}B(r)\Big]=\frac{r-r_s}{r}f(r)-\frac{r}{r-r_s}g(r),\nonumber\\
	&\kappa\Big[B'(r)-\frac{ r_s}{r(r-r_s)}B(r)\Big]=\frac{2(r-r_s)}{r}f(r),\nonumber\\
	&\kappa\Big[A'(r)+\frac{ r_s}{r(r-r_s)}A(r)\Big]=\frac{2r}{r-r_s}c(r),\nonumber\\
	&\kappa(r-r_s)A(r)=\frac{r}{r-r_s}e(r).\eea
	Of course we recover the previous case when $A(r)=0$ and $c(r)=e(r)=0$. Using the EoM in \eqref{3}, we can solve for the function determining the metric perturbation  $h_{11}$ in terms of the function determining the $h_{00}$ perturbation
	\bea A(r)&=\frac{r^2}{(2r-3r_s)(r-r_s)^2}\Big[r_s B(r)+r(r-r_s)\left(C_1r-2B'(r)\right)\Big].\eea
	%again if $A(r)=0$ we back to \eqref{B}.
	In this case, we have two components of the Schwarzschild metric that have been perturbed, and the total contributions to these components read
	\bea g_{00}&=\left(1-\frac{r_s}{r}\right)+\kappa B(r)\nonumber\\
	g_{11}&=-\frac{1}{1-\frac{r_s}{r}}-\kappa \frac{r^2}{(2r-3r_s)(r-r_s)^2}\Big[r_s B(r)+r(r-r_s)\left(C_1r-2B'(r)\right)\Big]\eea
	Finally, one can see that when considering more general perturbations for the connection and the graviton we find generically a displaced horizon whose value is given by
	\be r=r_s+\kappa r_s B(r_s).\ee
	The previous case corresponded to $A(r)=0$ for which we have $B(r_s)=0$ and the horizon is not perturbed.
\end{itemize}
%%%%%%%%%%%%%%%%%%%%%%%%%%%%%%%%%%%%%%%%%%%%%%%%%%%%%%%%%%%%%%%%%%%%%%%%%%%%%%%%%%%%%%%%%%%%%%%%%%%%%%%%%%%%%%%%%%%%%%%%%%%%%%%
\section{The fate of spherical horizons in the restricted first order approach }
%%%%%%%%%%%%%%%%%%%%%%%%%%%%%%%%%%%%%%%%%%%%%%%%%%%%%%%%%%%%%%%%%%%%%%%%%%%%%%%%%%%%%%%%%%%%%%%%%%%%%%%%%%%%%%%%
In this final section, we return to the initial action \eqref{action2} including the cosmological constant to carry out the same analysis of section 2 but in the restricted FO formalism. Let us recall that with restricted we mean that we take as the background connection the Levi-Civita connection. As we will see, even in the restricted case, FO and SO are not strictly equivalent. Our purpose here is precisely to contrast both approaches to the problem. We already know as a matter of fact that the Schwarzschild-de Sitter metric is an exact solution of the graviton EoM, but we want to see whether the other families that were studied in SO are also solutions in the restricted FO approach. 

%Before going into the details of the different families found, let us summarize the result in this FO approach.
Again with the ansatz \eqref{metricansatz} of the metric in the EoM \eqref{FOEOM}, and restricting ourselves to the background connection being the Levi-Civita one, we have that the EoM of the metric take the following form in terms of $A(r)$ and $B(r)$
\bea \label{EOMCFO}
&H^{FO}_{00}=\dfrac{1}{24 r^4A^4 B^3} \Big\{-4A^4 B^4\left(4\a-12\b-6r^2\g+3r^4\Lambda\right)+8A^3 B^3\left((4\a-12\b-3r^2\g)B+6r\a B^{'}\right)+\nonumber\\
&+r^2B^2A'^2\left(-4(\a-12\b)B^2-12r\a BB'+r^2(\a-3\b)B'^2\right)-2rABA'\Big[8(\a+6\b)B^3-r^3(\a-3\b)B'^3-\nonumber\\
&-12r\a B^2(B'+rB^{''})+2r^2 B B'\left((8\a-6\b)B'+r(\a-3\b)B''\right)\Big]+\nonumber\\
&+A^2\Big[8B^4\left(-2\a+6\b+r(2\a+12\b+3r^2\g)A'\right)-48r\a B^3B'+r^4(\a-3\b)B'^4-\nonumber\\
&-4r^3BB'^2\left((5\a-6\b)B'+r(\a-3\b)B''\right)+4r^2B^2\left((7\a-12\b)B'^2+2r(5\a-6\b)B'B''+\right.\nonumber\\
&\left.+r^2(\a-3\b)B''^2\right)\Big]\Big\}= 0,
\eea 
%%%%%%%%%%%%%%%%%%%%%%%%%%%%%%%%%%%%%%%%%%%%%%%%%%%%%%%%%%%%%%%%%%%%%%%%%%%%%%%%%%%%%%%%%%%%%%%%%%%%%%%%%%%%%%%%
\bea
\label{EOMDFO}
&H^{FO}_{11}= \dfrac{1}{24 r^4A^3 B^4} \Big\{4A^4 B^4\left(4\a-12\b-6r^2\g+3r^4\Lambda\right)+8A^3 B^3\left(-(4\a-12\b-3r^2\g)B+\right.\nonumber\\
&\left.+r(2\a+12\b+3r^2\g)B'\right)+r^2B^2A'^2\left((-28\a+48\b)B^2-4r(5\a-6\b) BB'-r^2(\a-3\b)B'^2\right)+\nonumber\\
&+2rABA'\Big[-24\a B^3-r^3(\a-3\b)B'^3+4rB^2(-3\a B'+r(5\a-6\b)B'')+\nonumber\\
&+2r^2 B B'\left((-8\a+6\b)B'+r(\a-3\b)B''\right)\Big]+\nonumber\\
&+A^2\Big[16B^4\left(\a-3\b+3r\a A'\right)-16r(\a+6\b) B^3B'-r^4(\a-3\b)B'^4+\nonumber\\
&+4r^3BB'^2\left(-3\a B'+r(\a-3\b)B''\right)-4r^2B^2\left(-(\a-12\b)B'^2-6r\a B'B''+\right.\nonumber\\
&\left.+r^2(\a-3\b)B''^2\right)\Big]\Big\}= 0,
\eea
%%%%%%%%%%%%%%%%%%%%%%%%%%%%%%%%%%%%%%%%%%%%%%%%%%%%%%%%%%%%%%%%%%%%%%%%%%%%%%%%%%%%%%%%%%%%%%%%%%%%%%%%%%%%%%%%
\bea
\label{EOMEFO}
&H^{FO}_{22}= \dfrac{1}{24 r^4A^4 B^4} \Big\{4 (-4 \a + 12 \b + 3 r^4 \Lambda) A^4 B^4+r^2 B^2 A'^2 (12 \a B^2 + 
4 r (\a - 3 \b) BB'+\nonumber\\
& + 
r^2 (\a - 3 \b) B'^2)+2 r A B A' \Big(8 (\a - 3 \b) B^3 + 
r^3 (\a - 3 \b)B'^3 - 
2 r^3 (\a - 3 \b) B B' B''-\nonumber\\
&- 
4 r B^2 (-3 \a B' + 
r (\a - 3 \b)B'')\Big)+2 A^3 B^2 \Big(16 (\a - 3 \b) B^2 - 
3 r^4 \g B'^2 + \nonumber\\
&+
B ((8 r (\a - 3 \b) + 6 r^3 \g)B' + 6 r^4 \g B'')\Big)+A^2 \Big[-4 B^4 (4 (\a - 3 \b) + 
r (4 \a - 12 \b + 3 r^2 \g) A') -\nonumber\\
&- 2 r B^3 (8 (\a - 3 \b) + 
3 r^3 \g A') B' + 
r^4 (\a - 3 \b) B'^4 - 
4 r^3 (\a - 3 \b) BB'^2 (B' + rB'')+ \nonumber\\
&+
4 r^2 B^2 (3 \a B'^2 + 
2 r (\a - 3 \b)B' B'' + 
r^2 (\a - 3 \b) B''^2)\Big]\Big\}= 0,
\eea
%%%%%%%%%%%%%%%%%%%%%%%%%%%%%%%%%%%%%%%%%%%%%%%%%%%%%%%%%%%%%%%%%%%%%%%%%%%%%%%%%%%%%%%%%%%%%%%%%%%%%%%%%%%%%%%%
and $H^{FO}_{33}=H^{FO}_{22}\sin^2\theta$, where $A^{(n)}$ corresponds to the n-th partial derivative. On the other hand, taking the EoM for the connection given in \eqref{FOEOM}, we have five independent components
\bea 
&C^1_{00}=\frac{1}{3 r^3 A^3 B^2} \Big\{-8 (2 \a + 3 \b) A^3 B^3 + 
2 r^2 B^2 A'^2 (-4 (2 \a + 3 \b) B
+ 
r (\a - 3 \b) B') + \nonumber\\
&
+r^2 A B \Big[2 r (\a - 3 \b)A' B'^2 + 
4 (2 \a + 3 \b) B^2 A'' - (\a - 3 \b) B (r B' A'' + 
A' (4B' + 
3 r B'')) \Big]+ \nonumber\\
&+
2 A^2 \Big[4 (2 \a + 3 \b) B^3 + 
r^3 (\a - 3 \b) B'^3 - 
2 r^2 (\a - 3 \b) BB'(B' + r B'') + \nonumber\\
&+
r (\a - 3 \b) B^2 (-2B' + 
r (2 B''+ r 
B^{(3)}))\Big]\Big\},
\nonumber \\
\nonumber \\
%%%%%%%%%%%%%%%%%%%%%%%%%%%%%%%%%%%%%%%%%%%%%%%%%%%%%%%%%%%%%%%%%%%%%%%%%%%%%%%%%%%%%%%%%%%%%%%%%%%%%%%%%%%%%%%%
&C^0_{10}=\frac{1}{6 r^3 A^3 B^2} \Big\{-8 (\a - 3 \b) A^3 B^3 - 
2 r^2 (\a - 3 \b) B^2 A'^2 (4 B + r B') + \nonumber\\
&+r^2 A B \Big[-2 r (\a - 3 \b) A' B'^2 + 
4 (\a - 3 \b) B^2 A'' + 
B\Big(r (\a- 3 \b) B' 
A'' - 
A' (4 (2 \a + 3 \b) B' -\nonumber\\
&- 
3 r (\a- 3 \b)B'')\Big)\Big]+
2 A^2 \Big[4 (\a - 3 \b) B^3 - 
r^3 (\a- 3 \b)B'^3 + \nonumber\\
&+
2 r^2 B B' (-(2 \a + 3 \b) B' + 
r (\a - 3 \b)B'')+ 
r (\a - 3 \b) B^2 (2 B' - r (2 B'' + r 
B^{(3)}))\Big]\Big\},
 \nonumber \\
\nonumber \\
%%%%%%%%%%%%%%%%%%%%%%%%%%%%%%%%%%%%%%%%%%%%%%%%%%%%%%%%%%%%%%%%%%%%%%%%%%%%%%%%%%%%%%%%%%%%%%%%%%%%%%%%%%%%%%%%
&C^1_{11}=\frac{1}{r^3 A B^2} \Big\{4 \a\Big[2 A^2 B^2 - 
r B A' (B + r B') - 
A (2 B^2 + r^2B'^2 + 
r B(B' - r B''))\Big]
\Big\},
 \nonumber \\
 \nonumber \\
%%%%%%%%%%%%%%%%%%%%%%%%%%%%%%%%%%%%%%%%%%%%%%%%%%%%%%%%%%%%%%%%%%%%%%%%%%%%%%%%%%%%%%%%%%%%%%%%%%%%%%%%%%%%%%%%
&C^2_{12}=\frac{1}{6 r A^3 B^3} \Big\{-8 (2 \a + 3 \b) A^3 B^3 + 
2 r^2 (\a - 3 \b) B^2 A'^2 (4 B + r B') +
r A B \Big[2 r^2 (\a - 3 \b) A' B'^2 + \nonumber\\
&+ 
4 B^2 (3 \a A' - 
r (\a - 3 \b)A'') + 
r B \Big(-r (\a - 3 \b) B'A''+ 
A' (2 (\a + 6 \b) B'- 
3 r (\a - 3 \b) B'')\Big)\Big]+\nonumber\\
& + 
2 A^2 \Big[4 (2 \a + 3 \b) B^3 + 
r^3 (\a - 3 \b) B'^3 +r^2 B B' ((\a + 6 \b) B' - 
2 r (\a - 3 \b) B'') +\nonumber\\
&+ 
r B^2 ((4 \a + 6 \b) B' + 
r (-2 (2 \a + 3 \b) B''+ 
r (\a - 3 \b) 
B^{(3)}))\Big]\Big\}, 
\eea
%%%%%%%%%%%%%%%%%%%%%%%%%%%%%%%%%%%%%%%%%%%%%%%%%%%%%%%%%%%%%%%%%%%%%%%%%%%%%%%%%%%%%%%%%%%%%%%%%%%%%%%%%%%%%%%%
\bea 
&C^1_{22}=\frac{1}{3 r A^3 B^3} \Big\{-8 (\a - 3 \b) A^3 B^3 + 
2 r^2 B^2 A'^2 (2 (\a + 6 \b) B + 
r (2 \a + 3 \b) B') -\nonumber\\
&- 
r A B \Big[-2 r^2 (2 \a+ 3 \b) A' B'^2 + 
2 B^2 (-3 \a A' + 
r (\a + 6 \b) A'') + \nonumber\\
&+
r B \Big(r (2 \a + 3 \b) B'A'' + 
A' (2 (\a + 6 \b)B'+ 
3 r (2 \a + 3 \b) B'')\Big) \Big]+\nonumber\\
&+ 
2 A^2 \Big[4 (\a - 3 \b) B^3 + 
r^3 (2 \a + 3 \b) B'^3 - 
r^2 B B' ((\a + 6 \b) B' + 
2 r (2 \a + 3 \b) B'') + \nonumber\\
&+
r B^2 (-(\a + 6 \b)B' + 
r ((\a + 6 \b) B'' + 
r (2 \a+ 3 \b) 
B^{(3)}))\Big]
\Big\}.\eea 
%%%%%%%%%%%%%%%%%%%%%%%%%%%%%%%%%%%%%%%%%%%%%%%%%%%%%%%%%%%%%%%%%%%%%%%%%%%%%%%%%%%%%%%%%%%%%%%%%%%%%%%%%%%%%%%%
\subsection{Solutions near the origin}
%%%%%%%%%%%%%%%%%%%%%%%%%%%%%%%%%%%%%%%%%%%%%%%%%%%%%%%%%%%%%%%%%%%%%%%%%%%%%%%%%%%%%%%%%%%%%%%%%%%%%%%%%%%%%%%%
\par
Let us try and find power series solutions as general as possible to the equations of motion presented above, mirroring what we did in section 2 in the SO approach. We continue to classify the solutions by the behavior of the functions appearing in the metric \eqref{metricansatz}  at $r\sim 0$, where we expand them as
\bea
&A(r)=a_s r^s+a_{s+1}r^{s+1}+\ldots,\nonumber\\
&B(r)=b_t\left(r^t+b_{t+1}r^{t+1}+\ldots\right).
\eea
Again, we classify the solutions in terms of the values of $(s,t)$.
\bi
\item There is a one 1-parameter, $\Lambda$, family with the behavior $(s,t)=(0,0)$. 
\bea &A(r)=1+\left(\frac{r}{r_\Lambda}\right)^2+O(r^4)\nonumber\\
&\frac{B(r)}{b_0}=1-\left(\frac{r}{r_\Lambda}\right)^2+O(r^4),
\eea
These simple solutions do not possess any singularity and correspond to candidates for the vacuum of the theory (of course, if $\Lambda=0$ this family reduces to Minkowski). In this case, we see that these solutions are fully characterized by the cosmological constant, that is, they correspond to constant curvature spacetimes whose properties will be driven by the sign of $\Lambda$. This family is nothing but the expansion of the de Sitter spacetime around the origin. 
\item There is another 2-parameter, $(a_1, \Lambda)$, singular family with the behavior $(s,t)=(1,-1)$. 
\bea 
&A(r)=a_1 r-a_1^2 r^2+a_1^3 r^3+\left(-a_1^4+\left(\frac{a_1}{r_\Lambda}\right)^2\right) r^4 + O(r^5)  \nonumber \\
&\frac{B(r)}{b_{-1}}=\frac{1}{r}+a_1-a_1\left(\frac{r}{r_\Lambda}\right)^2 +O(r^3)
\eea
these solutions matchs exactly the  expansion of the Schwarzschild-de Sitter metric for 
\be
a_1=-\frac{1}{r_s }, \quad b_{-1}=-r_s.
\ee
\ei
\par
The remarkable result here is that the horizonless $(2,2)$ family that was present in the SO formalism is not a solution in the restricted FO approach. We study this in detail in the following section. 

%%%%%%%%%%%%%%%%%%%%%%%%%%%%%%%%%%%%%%%%%%%%%%%%%%%%%%%%%%%%%%%%%%%%%%%%%%%%%%%%%%%%%%%%%%%%%%%%%%%%%%%%%%%%%%%%%%%%%%%%%%%%%%%
\subsection{Absence of the (2,2) family}
%%%%%%%%%%%%%%%%%%%%%%%%%%%%%%%%%%%%%%%%%%%%%%%%%%%%%%%%%%%%%%%%%%%%%%%%%%%%%%%%%%%%%%%%%%%%%%%%%%%%%%%%%%%%%%%%%%%%%%%%%%%%%%%
In second order (SO) we found a family of solutions with the behaviour $(s,t)=(2,2)$, where the metric functions took the form
\bea &A(r)=a_2 r^2+a_2b_3 r^3-r^4\frac{a_2}{6}\left(2a_2+b_3^2-8b_4\right)+a_5 r^5+ O(r^6),\nonumber\\
&\frac{B(r)}{b_2}=r^2+b_3 r^3+b_4 r^4+b_5 r^5+ O(r^6).
\eea
These solutions depended on six parameters $(b_3, b_4, b_5, a_2, a_5,\Lambda)$. This family is physically very interesting as it represents a horizonless family that can offer an alternative outcome of a spherical symmetric collapse in SO quadratic gravity. This family, however,  is not present anymore in the restricted FO approach. Let us now explore why this happens. Starting with  a general Lagrangian $\mathcal{L}(g,\Gamma)$, the second order equation of motion is, omitting indices in order not to clutter the notation, 
\be\label{SO}
\frac{\d\mathcal{L}}{\d g}+\frac{\mathcal{\d L}}{\d \Gamma}\frac{\d\Gamma(g)}{\d g}=0,
\ee
where we have used the chain rule as we know that $\Gamma_{SO}= \Gamma_{LC} = \Gamma(g)$.
In the FO approach, however, we perform independent variations for the connection and the metric field, although in the restricted case studied here we then fix the connection to be the Levi-Civita one. Nevertheless, we find the system of equations
\bea &\frac{\d\mathcal{L}}{\d g}=0\nonumber\\
&\frac{\mathcal{\d L}}{\d \Gamma}=0\eea
From here one can clearly see that all the solutions of FO formalism will be solutions in the SO one. Nevertheless, the converse is not true. It may happen that the two summands in \eqref{SO} cancel each other without neither of them vanishing. This is exactly what happens for the  $(s,t)=(2,2)$ SO solution as we can easily see below.
\par
We begin with the form of the $(2,2)$ family of solutions,
\bea &A(r)=a_2 r^2+ O(r^3),\nonumber\\
&B(r)=b_2r^2+ O(r^3).
\eea 
Introducing this in the EoM of the graviton \eqref{EOMFOG}, i.e. $\frac{\d\mathcal{L}}{\d g}=0$, we find
\bea
&H^{FO}_{00}=-192\a a_2^2 b_2^4=0\nonumber\\
&H^{FO}_{11}=-576\a a_2^2 b_2^4=0\eea
On the other hand, we can take the EoM for the connection field \eqref{EOMFOC} and contract it with the variation of the connection (the Levi-Civita one in the restricted approach) with respect to the metric function to compare it with the second summand of \eqref{SO}. We get
\bea
&D_{\m\n}=-2\a\left(\widebar{\nabla}_\m\widebar{\nabla}_\n\bR+2\widebar{R}_{\m\l}\widebar{R}_\n^{~\l}-2\bR^{\r\s}\bR_{\m\r\n\s}-\widebar{\Box}\bR_{\m\n}-\frac{1}{2}\bg_{\m\n}\widebar{\Box}\bR\right)+\nonumber\\
&+2\left(\b+\frac{2\a}{3}\right)\left(\widebar{\nabla}_\m\widebar{\nabla}_\n\bR-\bg_{\m\n}\widebar{\Box}\bR\right)=0.\eea
Introducing the ansatz of the $(2,2)$ family of solutions they read
\bea
&D_{00}=192\a a_2^2 b_2^4=0,\nonumber\\
&D_{11}=576\a a_2^2 b_2^4=0.
\eea 
In the restricted FO we have to solve them the the system of equations $H^{FO}_{00}=H^{FO}_{11}=0$ and $D_{00}=D_{11}=0$, whose solution is $a_2=0$ or $b_2=0$, i.e. there is no $(2,2)$ family. On the other hand in SO, the EoM \eqref{SO} is nothing but the sum given by $H^{SO}_{\m\n}=H^{FO}_{\m\n}+D_{\m\n}$, which one can trivially see that is fulfilled in this case, the $(2,2)$ family being a solution in SO. 

%%%%%%%%%%%%%%%%%%%%%%%%%%%%%%%%%%%%%%%%%%%%%%%%%%%%%%%%%%%%%%%%%%%%%%%%%%%%%%%%%%%%%%%%%%%%%%%%%%%%%%%%%%%%%%%%%%%%%%%%%%%%%%%
\section{Conclusions}
%%%%%%%%%%%%%%%%%%%%%%%%%%%%%%%%%%%%%%%%%%%%%%%%%%%%%%%%%%%%%%%%%%%%%%%%%%%%%%%%%%%%%%%%%%%%%%%%%%%%%%%%%%%%%%%%
Let us begin with the SO approach. First of all, we have generalized the analysis in \cite{Stelle78,Lu,Holdom,Holdom:2016nek}  to the case where a cosmological constant is present. It can be proven that the different families of solutions found in the references above still hold and possess the same behavior. In particular, we find that near the origin $r=0$, we still have the horizonless $(0,0)$ and $(2,2)$ families and the $(1,-1)$ Schwarzschild-de Sitter-like family. 
One of the main points of the paper is the fact that we can match these families of solutions (with different behavior near the origin) with the asymptotic universal behavior in the infinity, the $(-2,2)$ family.  
\par
From the physical point of view the most interesting question is whether the horizonless $(2,2)$ solution is compatible with Solar system tests of General Relativity; in other words, how big is the difference between it and the template Schwarzschild solution, or more generally, whether it qualifies as a possible candidate for the endpoint of stellar evolution in appropriate situations. This point has been recently studied by Holdom \cite{Holdom:2016nek,Holdom:2019ouz,Holdom:2020onl}. Physically, the region we can test is almost the asymptotic one, that is, $r\gg r_s$ although recent effort has been made regarding the possible experimental signatures of the horizon region (see cf. \cite{Cardoso:2016rao,Abedi:2016hgu}). The region in the vicinity of the singularity $r\sim 0$ is out of experimental reach for the time being. 
\par
Concerning that, the most important fact is that the asymptotic expansion in the regime appropriate for Solar system observations (namely $ {r\over r_s}\gg 1$ but ${r\over r_\Lambda}\ll 1$) is quite universal, in the sense that it is not much affected by the presence of higher dimension operators. It could even be said that it is structurally stable.
Thus, it seems that most of the observational evidence \cite{Will} supporting Schwarzschild's solution is also valid for the $(2,2)$ horizonless solutions. An important physical problem is then whether it is possible to find a way of telling between these horizonless solutions and the Schwarzschild one. This has already been attempted with some numerical analysis  \cite{Holdom:2016nek,Holdom:2019ouz,Holdom:2020onl,Cardoso:2016rao}. Gravitational wave probes regarding quadratic theories of gravity and the horizonless solutions have also been studied in \cite{Babourova:1998ct,Fonseca:2011aa,Wagle:2019mdq,Cardoso:2016oxy}. More work is clearly needed before this issue is sorted out.
\iffalse
%%%%%%%%%%%%%%%%%%%%%%%%%%%%%%%%%%%%%%%%%%%%%%%%%%%%%%%%%%%%%%%%%%%%%%%%%%%%%
\par 
We have also uncovered a region of parameter space in which the (2,2) solution is absent and only Einstein spaces are allowed. This corresponds to the analysis of the Riemannian minimization of the particular case of quadratic Lagrangians that can be written as a sum of squares. In this region of the parameter space, there is a constraint between the cosmological constant, Newton's constant, and the coupling constants of the quadratic invariants. Moreover, the positivity of the coefficients multiplying the sum of squares fixes the sign of the cosmological constant to be negative, that is, the solutions are of Schwarzschild-anti de Sitter type. 
%%%%%%%%%%%%%%%%%%%%%%%%%%%%%%%%%%%%%%%%%%%%%%%%%%%%%%%%%%%%%%%%%%%%%%%%%%%%%%%%
\fi
\par
Unfortunately, and even if we put our wagers in the SO formalism, no conclusions can be drawn from our analysis on the final state of celestial bodies, which is itself a dynamical process. In this paper, we have only analyzed the structural stability of solutions of Einstein's equations concerning small modifications of said equations by terms coming from contributions of the Lagrangian of higher-order in the curvature. In fact, only quadratic terms have been considered, but all orders should be taken into account for consistency because all orders will appear as counterterms in a perturbative quantum treatment.
\par
We would like not to be misinterpreted. We are not claiming that the $(2,2)$ solution is the only one that matches correctly with the asymptotic behavior. We are only claiming that it matches as well as the $(-1,1)$ one. Our guess is that which particular solution is the correct one depends on the physical situation at hand. There is no unique response; remember that Birkhoff's theorem does not hold anymore. The dynamics of a realistic collapse is an even more involved problem now than in the Einstein-Hilbert theory.
\par
In the second part of the paper, we have tackled the same problem in the FO approach. 
%To begin with, following what we have done in SO, only lower spin (0 and 2) spherically symmetric perturbations have been considered. 
In the particular cases where the background spacetime is Ricci flat, the graviton EoM is tautological (proportional to Lanczos' identity), so that the only spin 2 perturbation is to be found in one of the spin 2 components present in the connection field. We have worked out some simple examples in constant curvature spaces to check how this formalism works. 
\par
In  the general case, however, when the graviton EoM has got non-trivial dynamical content (this happens 
for the families of solutions studied in section 7) we follow the FO approach restricted to the Levi-Civita connection. We first point out that FO and SO are not equivalent in general even in this restricted case. Actually, we find that the horizonless $(2,2)$ power series solution of the SO equations is {\em not} a solution of the graviton EoM in the FO approach. This is a physically relevant manifestation of the non-equivalence of FO and SO approaches in the present context. It is quite remarkable that this non-equivalence appears with respect to one of the physically most interesting physical solutions.
\par
At any rate, and even if one were to believe the SO results on the existence of horizonless solutions, it is still possible, of course, that some unknown as yet dynamical law of nature prevents the disappearance of horizons. It could be, for example, that consistency forces an {\em unnatural} value for all renormalized  coefficients of the higher-order operators in the effective action, namely that all are to be set equal to zero. The only remaining operator will then be exactly $R$ so that the Einstein-Hilbert Lagrangian would be stable after renormalization. 
\par
It is of course also possible that the ultraviolet completion of general relativity is only possible in terms of other variables, such as strings. Nevertheless, even in this case, we believe our arguments to be sound.
\par
Precisely in that respect, the {\em Cosmic censorship hypothesis}  \cite{Penrose} has been advanced by Roger Penrose, conjecturing that all singularities should be veiled by a corresponding horizon. As a matter of fact, the physical mechanism hiding naked singularities remains largely unknown. At any rate, it should be remembered that predictivity is anyway lost at the {\em Big Bang}, which is in some sense the mother of all singularities.
%%%%%%%%%%%%%%%%%%%%%%%%%%%%%%%%%%%%%%%%%%%%%%%%%%%%%%%%%%%%%%%%%%%%%%%%%%%%%%%%%%%%%%%%%%%%%%%%%%%%%%%%%%%%%%%%%%%%%%%%%%%%%%
\section{Acknowledgements}
%%%%%%%%%%%%%%%%%%%%%%%%%%%%%%%%%%%%%%%%%%%%%%%%%%%%%%%%%%%%%%%%%%%%%%%%%%%%%%%%%%%%%%%%%
One of us (EA) acknowledges stimulating discussions with Gonzalo Olmo and Tom\'as Ort\'{\i}n. This work has received funding from the Spanish Research Agency (Agencia Estatal de Investigacion) through the grant IFT Centro de Excelencia Severo Ochoa SEV-2016-0597, and the European Union's Horizon 2020 research and innovation programme under the Marie Sklodowska-Curie grants agreement No 674896 and No 690575. We have also been partially supported by FPA2016-78645-P(Spain). RSG is supported by the Spanish FPU Grant No FPU16/01595. This project has received funding/support from the European Union Horizon 2020 research and innovation programme under the Marie Sklodowska-Curie grant agreement  860881-HIDDeN.

%%%%%%%%%%%%%%%%%%%%%%%%%%%%%%%%%%%%%%%%%%%%%%%%%%%%%%%%%%%%%%%%%%%%%%%%%%%%%%%%%%%%%%%%%%%%%%%%%%%%%%%%%%%%%%%%%%%%%%%%%%%%%%%
\newpage
\appendix
%%%%%%%%%%%%%%%%%%%%%%%%%%%%%%%%%%%%%%%%%%%%%%%%%%%%%%%%%%%%%%%%%%%%%%%%%%%%%%%%%%%%%%%%%%%%%%%%%%%%%%%%%%%%%%%%%%%%%%%%%%%%%%%
\section{Runaways and higher derivatives} \label{appB}
%%%%%%%%%%%%%%%%%%%%%%%%%%%%%%%%%%%%%%%%%%%%%%%%%%%%%%%%%%%%%%%%%%%%%%%%%%%%%%%%%%%%%%%%%%%%%%%%%%%%%%%%%%%%%%%%%%%%%%%%%%%%%
The archetypal physical problem in which higher derivatives appear is the back reaction of the electromagnetic radiation; id est, the Lorentz-Dirac equation \cite{Coleman}. Equations of motion of degree higher than two typically get runaway solutions. This is a purely classical phenomenon, which is related to the presence of ghosts when quantizing the system, but one that can be analyzed independently. Consider for simplicity a linear equation of third-degree in time derivatives
\be
{d^3\over dt^3}\, x(t)+a\,{d^2\over dt^2}\, x(t)+b \,{d \over dt} x(t)+ c\, x(t)=f(t).
\ee
this equation can be easily solved using Fourier transform. Nevertheless, it is quite easy to prove that there is always at least one solution of the form
\be
x_{run}=e^{\l t}
\ee
with real $\l$ (the sign depends on the details of the equation). This follows from a classical theorem that asserts that a cubic algebraic equation has got at least one real solution (it can have all three roots real, depending on the sign of the discriminant). These are the {\em runaway solutions}. Depending on the sign of the exponent, they grow in time without bound, or else, ${1/x}$ does it; they are not oscillatory solutions. As has been already indicated, they are the classical counterpart of quantum ghosts.
\par
It is a quite widespread belief that higher derivatives are always a problem \cite{AAGM,Alvarez-Gaume,Rham,Salvio} even when it naively appears that there are no ghosts or tachyons. Nevertheless, this should be qualified in some cases, as we shall see. A standard argument for the iterated d'Alembertian is as follows. Consider the Lagrangian \cite{Brust:2016gjy}
\be
L=\phi \Box^2\phi.
\ee
introducing a Lagrange multiplier
\be
L=\psi \Box^2 \phi-{1\over 4}\psi^2,
\ee
and making the field redefinitions 
\bea
&\psi=\phi_++\phi_-\nonumber\\
&\phi=\phi_+-\phi_-,
\eea
the Lagrangian takes the form
\be
L=\phi_+\Box^2\phi_+-\phi_-\Box^2\phi_--{1\over 4}(\phi_-+\phi_-)^2.
\ee
We then see that $\phi_-$ is a ghost because it has the wrong sign in its kinetic term. This argument is not very convincing though, because the fields $\phi_\pm$ are not independent.
\par
In spite of the fact that the delta function with support on the light cone
\be
\d((k^2)^2)
\ee
is not well-defined apriori, in the classical reference \cite{Schwartz}, it is shown that the equation
\be
\Box^2 \phi=0
\ee
has a well-defined Cauchy problem; that is, there is a unique solution determined by the initial conditions. This fact is a consequence of  the existence of  Riesz' distribution $G_\l (x)$ (to be explicitly defined in a moment) such that
\be
\Box^2 \lim_{\l\rightarrow 2}\, G_\l^+(x)=\d(x).
\ee
This, in turn, stems from  the fact that
\be
\Box G_\l=G_{\l-1},
\ee
as well as
\be
\lim_{\l\rightarrow 0} G_\l(x)=\d(x).
\ee
%\be
%\Box G^+(x)=\lim_{\l\rightarrow 1}    \Box G_\l^+(x)=G_0^+(x)=\d(x).
%\ee
We can write $G_\l^+$ as \cite{Friedlander} 
\be
G_\l^+(x)\equiv C_{n,\l}\,\g^{\l- n/2},
\ee
provided $x\in D^+(0)$ (the future domain of dependence of the origin) as $G_\l^+(x)$ vanishes otherwise. The quantity $\g(x)\equiv \Gamma(0,x)$ is Synge's world function from the origin to the point $x$, and the constant reads
\be
C_{n,\l}\equiv{1\over \pi^{\frac{n}{2}-1} 2^{2\l-1}(\l-1)!(\l- n/2)!}.
\ee
\par
In \cite{Brust:2016gjy} it is pointed out that even in those free theories, the Hilbert space of states lacks a positive definite scalar product so that there are indeed zero as well as negative norm states in it, that is, ghosts.
\par
All of this changes, of course, when more derivatives enter into the equation.
For example, in the particular case
\be
\left(\Box^2-2 m^2 \Box+m^4\right)\,\Phi=\left(\Box-m^2\right)^2\,\Phi=0,
\ee
there are runaway solutions 
\be
\Phi\sim e^{m t}.
\ee
This is exactly what happens in quadratic gravity. Making the ansatz
\bea\label{ASBS}
&A(r)=a e^{C_a r}\nonumber\\
&B(r)=b e^{C_b r},
\eea 
the trace of the equation of motion \eqref{EOMB} reads
\be H^\m_\m=-\g R-6\b{\Box} R-2\Lambda=0,\ee
and substituting the divergent ansantz \eqref{ASBS} we get
\bea &H^\m_\m=\frac{1}{2r^4 a^2}\Big\{-4a^2 r^2(-\g+r^2\Lambda)-\nonumber\\
&-e^{-rC_a}a\Big[48\b+4r^2\g+4(-6r\b+r^3\g)C_b+r^4\g C_b^2-C_a(-24r\b+4r^3\g+r^4\g C_b)\Big]-\nonumber\\
&-3e^{-2rC_a}\b\Big[3r^3C_a^3(4+rC_b)+4(-4+2rC_b+r^2C_b^2)-4r^2 C_a^2(2+5rC_b+r^2C_b^2)+\nonumber\\
&+rC_a(-24-4rC_b+8r^2 C_b^2+r^3C_b^3)\Big]\Big\}=0.\eea
This trace happens to be divergent whenever $C_a <0$.
Demanding cancellation of the more divergent terms (those that go as $e^{-2 r C_a}$)
we get the two possibilities
%\textcolor{red}{ESTA PARTE NO SE ENTIENDE}
%The solution is divergent if $C_a<0$, therefore the plus divergent terms are with $r^4$, i.e.
%\bea
%&&H^\m_\m\sim -\frac{3\b}{2a^2}e^{-2rC_a}C_aC_b\left(3C_a^2-4C_aC_b+C_b^2\right)
%\eea
%if we want to cancel this term
\be
C_b=\frac{1}{C_a} \quad \quad \text{or} \quad \quad C_b=\frac{1}{3C_a}.
\ee
As has been already mentioned in the main text, it would be important to isolate boundary (or initial) conditions that prevent those runaway solutions to appear. As far as we understand, this remains an outstanding problem very much worth exploring.

%%%%%%%%%%%%%%%%%%%%%%%%%%%%%%%%%%%%%%%%%%%%%%%%%%%%%%%%%%%%%%%%%%%%%%%%%%%%%%%%%%%%%%%%%%%%%%%%%%%%%%%%%%%%%%%%%%%%%%%%%%%%%%%
\section{Notes about constant curvature spaces (CCS)}\label{C}
%%%%%%%%%%%%%%%%%%%%%%%%%%%%%%%%%%%%%%%%%%%%%%%%%%%%%%%%%%%%%%%%%%%%%%%%%%%%%%%%%%%%%%%%%%%%%%%%%%
Synge's \cite{Synge} world function $\s(x,y)$ is defined as the square of the geodesic distance. It is appealing to use it  in pseudo Riemannian spaces as it is positive semidefinite. In the Riemannian case it is simpler to use directly the geodesic length, $s(x,\xp)$,  defined via the equation
\be 
\bg^{\m\n}~\pd_\m s ~\pd_\n s\equiv  s^\m~ s_\m= 1
\ee
It also simple to check that for the CCS with scale L which are the ones obeying
\be 
\bR_{\a\b\g\d}=\mp { 1\over
	L^2}\left(\bg_{\a\g}\bg_{\b\d}-\bg_{\a\d}\bg_{\b\g}\right),
\ee
then in the elliptic case (negative cosmological constant, positive scalar curvature, with our conventions)  the following formulas are true (the hyperbolic case which corresponds to positive cosmological constant and  negative scalar curvature is quite similar, with circular functions replaced by hyperbolic ones)
\bea\label{P}
&s_{\m\n}=s_{\n\m}={1\over L\tan~{s\over
		L}}\left(\bg_{\m\n}-s_\m s_\n\right), \nonumber\\
&s_{\m\n}s^\m=s_{\m\n}s^\m=0, \nonumber\\
&\bar{\Box} s={n-1\over L~\tan~{s\over L}}, \nonumber\\
&s_{\m\n\r}=-{1\over L^2\sin^2~{s\over
		L}}~s_\r \bg_{\m\n}-{1\over L^2~\tan^2~{s\over
		L}}\left(\bg_{\m\r}s_\n+s_\m \bg_{\n\r}\right)+{1+2~\cos^2~{s\over
		L}\over L^2~\sin^2~{s\over L}}~s_\m s_\n s_\r,\nonumber\\
&s_{\a\b\g}-s_{\a\g\b}=\bR_{\g\b\a\l}s^\l=\frac{\bR}{n(n-1)}\left(\bg_{\a\g}s_\b-\bg_{\a\b}s_\g\right),\nonumber\\
&\bar{\Box}(s_\m)=-{n-1\over
	L^2~\tan^2~{s\over L}}~s_\m, \nonumber\\
&s_{\a\l}~s^{\l}_{~\b}={1\over L^2~\tan^2~{s\over
		L}}\left(\bg_{\a\b}-s_\a s_\b\right), \nonumber\\
&s^\l s_{\a\b\l} =-{1\over L^2\sin^2~{s\over L}}~
\left(\bg_{\a\b}-s_\a s_\b\right),\nonumber\\
&s^\l_{~\l\m}\equiv(\bar{\Box}
s)_\m=-{n-1\over L^2~\sin^2~{s\over L}}~s_\m.
\eea
Some of these formulas are used in the main text.

\newpage
%%%%%%%%%%%%%%%%%%%%%%%%%%%%%%%%%%%%%%%%%%%%%%%%%%%%%%%%%%%%%%%%%%%%%%%%%%%%%%%%%%%%%%%%%%%%%%%%%%%%%%%%%%%%%%%%%%%%%%%%%%%%%%%

\end{document}